\documentclass[twocolumn,aps]{revtex4}

\usepackage[english]{babel}

\usepackage{amsmath}
\usepackage{graphicx}
\usepackage{multirow}
\usepackage{txfonts}
\def\bea{\begin{eqnarray}}
\def\eea{\end{eqnarray}}
\def\be{\begin{equation}}
\def\ee{\end{equation}}

\begin{document}

\title{Test of the cosmic evolution using Gaussian processes}

\author{Ming-Jian Zhang$^{b}$}
\author{Jun-Qing Xia$^{a}$}\email[Corresponding author: ]{xiajq@bnu.edu.cn}
\affiliation{$^a$Department of Astronomy, Beijing Normal University, Beijing 100875, China}
\affiliation{$^b$Key Laboratory of Particle Astrophysics, Institute of High Energy Physics,
Chinese Academy of Science, P. O. Box 918-3, Beijing 100049, China}

\begin{abstract}
Much focus was on the possible slowing down of cosmic acceleration under the dark energy parametrization. In the present paper, we investigate this subject using the Gaussian processes (GP), without resorting to a particular template of dark energy. The reconstruction is carried out by abundant data including luminosity distance from Union2, Union2.1 compilation and gamma-ray burst, and dynamical Hubble parameter. It suggests that slowing down of cosmic acceleration cannot be presented within 95\% C.L., in considering the influence of spatial curvature and Hubble constant. In order to reveal the reason of tension between our reconstruction and previous parametrization constraint for Union2 data, we compare them and find that slowing down of acceleration in some parametrization is only a ``mirage". Although these parameterizations fits well with the observational data, their tension can be revealed by high order derivative of distance $D$. Instead, GP method is able to faithfully model the cosmic expansion history.

\end{abstract}

\maketitle

\section{Introduction}
\label{introduction}

Multiple experiments have consistently confirmed the cosmic late-time accelerating expansion. Observations contributing to this pioneering discovery contain the type Ia supernova (SNIa) \citep{riess1998supernova,perlmutter1999measurements}, large scale structure \citep{tegmark2004cosmological}, cosmic microwave background (CMB)
anisotropies  \citep{spergel2003wmap}, and baryon
acoustic oscillation (BAO) peaks \citep{eisenstein2005detection}. Meanwhile, a large set of theoretical paradigms try to map this discovery, including the exotic dark energy with repulsive gravity, or modification to general relativity \citep{barrow1988inflation,dvali20004d}, or violation of cosmological principle \citep{lemaitre1933univers,tolman1934effect,bondi1947spherically}. In the dark energy doctrine, the cosmological constant model with an equation of state (EoS) $w=-1$ is the most notable candidate. Because of the lack of a clear theoretical understanding on the nature of dark energy, a wide variety of alternative models were developed. In particular, the Chevallier-Polarski-Linder (CPL) \cite{chevallier2001accelerating,linder2003exploring} model is also a potential competitor. Besides the dynamical theory, kinematics is another way to return the cosmic acceleration. Specifically, deceleration parameter $q(z)<0$ is a direct symbol of the cosmic acceleration. Conversely, $q(z) > 0$ implies the decelerating expansion. As well as the EoS, deceleration parameter is also unmeasurable and should be estimated by an ansatz, such as the linear form $q(z)=q_0 +q_1 z$ \cite{riess2004type}, or a nonlinear form $q(z)=q_0 + q_1 z/(1+z)$ \cite{xu2008constraints} etc.

Recently, a model conflicting with the mainstream research has caused wide public concern. In the drive of CPL model, Shafieloo et al. \cite{shafieloo2009cosmic} found that cosmic acceleration may have already peaked and that we are witnessing its slowing down at $z \lesssim 0.3$. In other words, they believed that current deceleration parameter may be $q_0 >0$ within 95\% C.L. Using the SNIa, BAO, CMB and lookback time data,  \citet{li2010probing,li2011examining} presented an extended study under three parameterized EoS including the CPL. They found that slowing down of acceleration was favored by Union2 SNIa compilation in CPL model, but it cannot be held in other considered models. Moreover, there exists a tension between the low redshift and high redshift data, especially adding the CMB data. In Refs. \cite{cardenas2012role,2013PhRvD..87d3502L}, the authors studied the CPL model again in the attendance of spatial curvature using different SNIa samples, BAO and CMB data. They found that incorporation of the spatial curvature can ameliorate above tension. However, it was still impossible to reach a consensus among the three types of data. Interestingly, a similar study on the CPL model found that slowing down of cosmic acceleration is also supported by the 42 measurements of the X-ray gas mass fraction $f_{gas}$ in clusters \cite{cardenas2013cosmic}. In Ref. \cite{vargas2012perturbations}, this was also supported in an interaction model between dark matter and dark energy. Recently, two comprehensive studies were made \cite{magana2014cosmic,wang2016comprehensive}. In these systematic researches, more parameterizations and larger volume of observational data were adopted. They both believed that slowing down of acceleration is a theoretical possibility which cannot be convincingly recognized by current data. Moreover, we noted that most of the investigations were performed in the specific dark energy model. The problem, subsequently, is how to break the limitation from dark energy parametrization, and to identify this conjecture with high accuracy.

We need a model-independent analysis on this interesting subject. The reason is that analysis driven by a strong dependence on the dark energy pattern, such as the CPL model, maybe fall our extrapolation. In the present paper, we will adopt the Gaussian processes (GP), a powerful model-independent technique, to provide a systematic analysis on the deceleration parameter. Unlike the parametrization constraint, this approach does not rely on any artificial dark energy template. It depends on the covariance function which can be estimated by the observational data. It is thus able to faithfully model the deceleration parameter at different redshift locations. In cosmology, it has incurred a wide application in reconstructing dark energy \cite{holsclaw2010nonparametric,seikel2012reconstruction} and cosmography \cite{shafieloo2012gaussian}, or testing standard concordance model \cite{yahya2014null} and distance duality relation \cite{santos2015two}, or determinating the interaction between dark matter and energy \cite{yang2015reconstructing} and spatial curvature \cite{cai2016null}. In the pedagogical introduction to GP, \citet{seikel2012reconstruction} invented the public code GaPP (Gaussian Processes in Python). They investigated the deceleration parameter only using SNIa data in the flat universe. In the present paper, we extend relevant study to a more general universe, using the SNIa, gamma-ray burst (GRB), and the dynamical measurement from observational $H(z)$ data (OHD). By the way, GRB cannot be directly accepted as good distance indicators. This is because of the lack of a set of low-redshift GRBs which are cosmology-independent. Here we use ``the calibrated" GRBs in a cosmology-independent method.

Our goal in the present paper, on the one hand, is to present a model-independent analysis on the cosmic evolution by deceleration parameter. On the other hand, we develop the relevant analysis using larger volume observational data with higher redshift, such as the distance measurement from GRB. Moreover, we also intend to investigate the impact of spatial curvature and Hubble constant $H_0$. In Section~\ref{kinematics}, we prepare ourselves to introduce the cosmic kinematics. And in Section~\ref{methodology} we introduce the GP approach and relevant data. We present the reconstruction result in Section \ref{result}. In Section \ref{Union2_discussion} we compare our result with previous parametrization constraint for the Union2 data, and try to find the reason of tension between them. Finally, in Section \ref{conclusion} conclusion and discussion are drawn. Throughout the paper, we use the natural units of the speed of
light $c=1$.

\section{Cosmic kinematics}
\label{kinematics}

Before the introduction of GP method, we should prepare ourselves in the cosmic kinematics. Different from the dynamical cosmological model, it does not need any assumption about the component of the universe, including the spatial curvature. In the FRW framework, the distance modulus of SNIa, difference between the apparent magnitude $m$ and the absolute magnitude $M$ can be estimated as
    \begin{equation}  \label{mu:define}
    \mu (z) = m(z) - M = 5 \textrm{log}_{10}d_L(z)+25,
    \end{equation}
where the luminosity distance function $d_L(z)$  is
\begin{equation}
    \label{dL:define}
    d_L(z) =  \frac{1 + z}{\sqrt{\left|\Omega_k \right|}}
    \operatorname{sinn} \left[\sqrt{\left|\Omega_k \right|} \int^z_0 \frac{ \mathrm{d}
    z'}{H(z')} \right],
\end{equation}
in which the $\operatorname{sinn}$ function therein is a shorthand for
the definition
\begin{eqnarray}
    \operatorname{sinn}(x) =
    \begin{cases}
    \sinh x, & \Omega_k > 0, \\
          x, & \Omega_k = 0, \\
     \sin x, & \Omega_k < 0.
   \end{cases}
\end{eqnarray}
For convenience, we define a dimensionless comoving luminosity distance
\be   \label{D:define}
D(z) \equiv H_0 (1+z)^{-1} d_L (z) .
\ee
Therefore, substituting Eq. \eqref{D:define} into Eq. \eqref{mu:define}, we can obtain the distance $D(z)$ from the observational distance modulus.
Combing Eq. \eqref{D:define} and \eqref{dL:define}, taking derivative with respect to redshift $z$, we obtain the Hubble parameter with spatial curvature
\be  \label{Hz:define}
H(z) = \frac{H_0 \sqrt{1+\Omega_k D^2}}{D'} ,
\ee
where the prime denotes the derivative with respect to redshift $z$. In previous work, the space curvature is commonly ignored for simplicity. In following section, we will investigate the effect of spatial curvature on the reconstruction.

\begin{figure}
    \begin{center}
\includegraphics[width=0.35\textwidth]{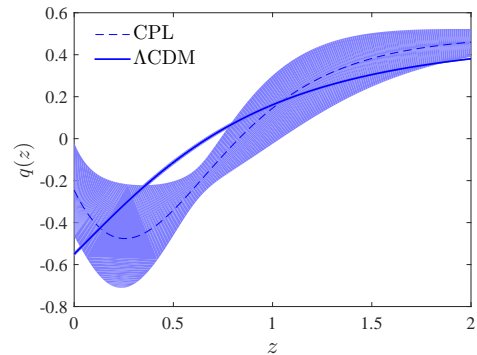}
    \end{center}
    \caption{\label{example} Toy model for the slowing down of cosmic acceleration in the $\Lambda$CDM model with $\Omega_m=0.3 \pm 0.01$ and CPL model with $w_0=-0.705^{+0.207}_{-0.212}$ and $w_a=-2.286^{+1.675}_{-1.469}$ \cite{wang2016comprehensive}.}
\end{figure}

In combination the relation
\begin{equation}  \label{qH:define}
   q(z)=  \frac{1+z}{H} \frac{\textmd{d} H}{\textmd{d} z} -1 ,
\end{equation}
with Eq. \eqref{Hz:define}, the deceleration parameter also can be expressed over the distance and spatial curvature
\be  \label{qD:define}
q(z) = \frac{\Omega_k D D'^2-(1+\Omega_k D^2)D''}{D'(1+\Omega_k D^2)} (1+z) -1 .
\ee
With Eq. \eqref{qH:define} and \eqref{qD:define}, we can reconstruct the deceleration parameter via OHD and distance measurements, respectively. In Figure \ref{example}, we plot the deceleration parameter in the fiducial $\Lambda$CDM model with $\Omega_m=0.3 \pm 0.01$  and CPL model with $w=w_0 + w_a z/(1+z)$, where $w_0=-0.705^{+0.207}_{-0.212}$ and $w_a=-2.286^{+1.675}_{-1.469}$ from recent literature \cite{wang2016comprehensive}. In common picture, $q(z)$ decreases from positive to negative with the decreasing redshift. However, for the popular CPL model, we note that $q(z)$ reaches its peak, and then eventually increase to a positive value within 95\% C.L., as shown in Figure \ref{example}. Certainly, bigger error of $w_a$ also influence the reconstruction, according to the error propagation. But we also note that best-fit $q(z)$ also reaches its trough at redshift $z \sim 0.25$. This is just the slowing down of cosmic acceleration widely discussed in previous work. In the following, we investigate it via the GP method.

\section{Methodology}
\label{methodology}

In this section, we report the related reconstructing method and observational data. To map the deceleration parameter, we need the $H(z)$ and its derivative in Eq. \eqref{qH:define}, and $D(z)$ and its derivatives in Eq. \eqref{qD:define}.

\subsection{Gaussian processes}  \label{GP}

In the parametrization constraint, a prior on the constrained function $f(z)$ is usually imposed, such as the CPL model with two artificial parameters $w_0$ and $w_a$. Instead, in the Gaussian processes, any parametrization assumption on the goal function $f(z)$ is redundant. Its key ingredient is the covariance function $k(z, \tilde{z})$ which correlates the function $f(z)$ at different points. Commonly, the covariance function $k(z, \tilde{z})$ with two hyperparameters $\ell$ and $\sigma_f$ can be determined by the observational data. Therefore, its model-independence promotes its wide application in the reconstruction of dark energy EoS \cite{holsclaw2010nonparametric1,holsclaw2010nonparametric,seikel2012reconstruction}, or in the test of the concordance model \cite{seikel2012using,yahya2014null}, or determination to the dynamics of dark energy by dodging the matter degeneracy \cite{busti2016dodging}.

For the covariance function $k(z, \tilde{z})$, many templates are available. The usual choice is the squared exponential $k(z, \tilde{z}) = \sigma_f^2 \exp[-|z-\tilde{z}|^2 / (2 \ell^2)]$.  Analysis in Ref. \cite{seikel2013optimising} shows that the Mat\'ern ($\nu=9/2$) covariance function is a better choice to present suitable and stable result. It thus has been widely used in previous work \cite{yahya2014null,yang2015reconstructing}. It is read as
\begin{eqnarray}
k(z,\tilde{z}) &=& \sigma_f^2
  \exp\Big(-\frac{3\,|z-\tilde{z}|}{\ell}\Big) \nonumber \\
  &&~\times \Big[1 +
  \frac{3\,|z-\tilde{z}|}{\ell} + \frac{27(z-\tilde{z})^2}{7\ell^2}  \nonumber\\
&&~~~~~~
+ \frac{18\,|z-\tilde{z}|^3}{7\ell^3} +
  \frac{27(z-\tilde{z})^4}{35\ell^4} \Big]. \label{mat}
\end{eqnarray}
With the chosen Mat\'ern ($\nu=9/2$) covariance function, we can reconstruct the deceleration parameter using the publicly available package GaPP developed by \citet{seikel2012reconstruction}. It also has been frequently used in above referenced work.

\subsection{Mock data}
\label{mock}

Following previous work, we demonstrate the ability of GP method first using some mock data. Our task here is to test whether the reconstruction agrees with the fiducial model and can be distinguished from the other fiducial model.

For the distance measurement, we create some simulated data via the Wide-Field InfraRed Survey Telescope-Astrophysics Focused Telescope Assets (WFIRST-AFTA) \footnote{http://wfirst.gsfc.nasa.gov/} in the fiducial flat $\Lambda$CDM model with $\Omega_m=0.3$. To distinguish from the $\Lambda$CDM model, we also perform the simulation in a CPL model with $w_0=-0.705$ and $w_a=-2.286$. The WFIRST-2.4 not only stores tremendous potential on some key scientific program, but also enables a survey with more supernova in a more uniform
redshift distribution. One of its science drivers is to measure the cosmic expansion history. According to the updated report by Science Definition Team \cite{spergel2015wide}, we obtain 2725 SNIa over the region $0.1<z<1.7$ with a bin $\Delta z=0.1$ of the redshift.
The photometric measurement error per supernova is $\sigma_{\textmd{meas}} = 0.08$ magnitudes. The
intrinsic dispersion in luminosity is assumed as $\sigma_{\textmd{int}} = 0.09$ magnitudes (after correction/matching for light curve shape and spectral properties). The other contribution to statistical errors is gravitational lensing magnification, $\sigma_{\textmd{lens}} = 0.07 \times z$
mags. The overall statistical error in each redshift bin is then
\be
\sigma_{\textmd{stat}} = \left[(\sigma_{\textmd{meas}})^2 + (\sigma_{\textmd{int}})^2 + (\sigma_{\textmd{lens}})^2 \right]^{1/2} / \sqrt{N_i} ,
\ee
where $N_i$ is the number of supernova in the $i$-th redshift bin. According to being
estimated, a systematic error per bin is
\be
\sigma_{\textmd{sys}} = 0.01 (1+z) /1.8  .
\ee
Therefore, the total error per redshift bin is
\be
\sigma_{\textmd{tot}} = \left[(\sigma_{\textmd{stat}})^2 + (\sigma_{\textmd{sys}})^2 \right]^{1/2} .
\ee

The other data we use is the Hubble parameter. Different from the distance measurement, a tailor-made program to measure $H(z)$ is not available.  \citet{ma2011power} evaluated the constraint power of SNIa and OHD, and found that as many as 64 further independent measurements of $H(z)$ are needed to match the constraining power of SNIa. Analysis to the error distribution of OHD, the simulated $H(z)$ can be empirically extrapolated to redshift $z<2$
\be
H_{\textmd{sim}} (z) = H_{\textmd{fid}} (z) + \Delta H ,
\ee
where $H_{\textmd{fid}} (z)$ is Hubble parameter in the fiducial models. The deviation $\Delta H$ is a random variable satisfying Gaussian distribution $N(0, \tilde{\sigma} (z))$. Empirically, the random number $\tilde{\sigma} (z)$ is also from a Gaussian distribution $N(\sigma_0 (z),  \varepsilon(z))$, where $\sigma_0 (z)=10.64 z +8.86$ is the mean uncertainty, $\varepsilon (z)= (\sigma_+ - \sigma_-)/4$ with $\sigma_+ = 16.87 z +10.84$ and $\sigma_- = 4.41 z +7.25$. With the mock data, we can safely test the reliability of reconstruction using GaPP. In following Section \ref{result}, we will seriously report the reconstruction results.

\begin{table}[tbp]
\caption{\label{tab: Hz} The current available OHD. The $H(z)$ and its uncertainty $\sigma_{H}$ are in the unit of km s$^{-1}$ Mpc$^{-1}$.  The upper panel are 30 samples deduced from the differential age method. The lower panel are 10 samples obtained from the radial BAO method. }
\centering
\begin{tabular}{cccc|cccc}
\hline
$z$&~~$H(z)$ &~~~$\sigma_{H}$&~Ref. & $z$&~~$H(z)$ &~~~$\sigma_{H}$&~Ref. \\
\hline\hline
$0.070$&~$69$&~~$19.6$&~\cite{zhang2014four} & $0.4783$&~$80.9$&~~$9$&~\cite{moresco2016A} \\
$0.090$&~$69$&~~$12$&~\cite{jimenez2003constraints} & $0.480$&~$97$&~~$62$&~\cite{stern2010cosmic} \\
$0.120$&~$68.6$&~~$26.2$&~\cite{zhang2014four} & $0.593$&~$104$&~~$13$&~\cite{moresco2012improved} \\
$0.170$&~$83$&~~$8$&~\cite{simon2005constraints} & $0.680$&~$92$&~~$8$&~\cite{moresco2012improved} \\
$0.179$&~$75$&~~$4$&~\cite{moresco2012improved} & $0.781$&~$105$&~~$12$&~\cite{moresco2012improved} \\
$0.199$&~$75$&~~$5$&~\cite{moresco2012improved} & $0.875$&~$125$&~~$17$&~\cite{moresco2012improved} \\
$0.200$&~$72.9$&~~$29.6$&~\cite{zhang2014four} & $0.880$&~$90$&~~$40$&~\cite{stern2010cosmic} \\
$0.270$&~$77$&~~$14$&~\cite{simon2005constraints} & $0.900$&~$117$&~~$23$&~\cite{simon2005constraints} \\
$0.280$&~$88.8$&~~$36.6$&~\cite{zhang2014four}& $1.037$&~$154$&~~$20$&~\cite{moresco2012improved} \\
$0.352$&~$83$&~~$14$&~\cite{moresco2012improved} & $1.300$&~$168$&~~$17$&~\cite{simon2005constraints} \\
$0.3802$&~$83$&~~$13.5$&~\cite{moresco2016A}&  $1.363$&~$160$&~~$33.6$&~\cite{moresco2015raising} \\
$0.400$&~$95$&~~$17$&~\cite{simon2005constraints}&  $1.430$&~$177$&~~$18$&~\cite{simon2005constraints} \\
$0.4004$&~$77$&~~$10.2$&~\cite{moresco2016A} & $1.530$&~$140$&~~$14$&~\cite{simon2005constraints} \\
$0.4247$&~$87.1$&~~$11.2$&~\cite{moresco2016A}&  $1.750$&~$202$&~~$40$&~\cite{simon2005constraints} \\
$0.44497$&~$92.8$&~~$12.9$&~\cite{moresco2016A} & $1.965$&~$186.5$&~~$50.4$&~\cite{moresco2015raising} \\
\\
$0.24$&~$79.69$&~~$2.65$&~\cite{gaztanaga2009clustering}&  $0.60$&~$87.9$&~~$6.1$&~\cite{blake2012wigglez} \\
$0.35$&~$84.4$&~~$7$&~\cite{xu2013measuring} & $0.73$&~$97.3$&~~$7.0$&~\cite{blake2012wigglez} \\
$0.43$&~$86.45$&~~$3.68$&~\cite{gaztanaga2009clustering} & $2.30$&~$224$&~~$8$&~\cite{delubac2013baryon} \\
$0.44$&~$82.6$&~~$7.8$&~\cite{blake2012wigglez} & $2.34$&~$222$&~~$7$&~\cite{delubac2015baryon} \\
$0.57$&~$92.4$&~~$4.5$&~\cite{samushia2013clustering} & $2.36$&~$226$&~~$8$&~\cite{font2014quasar} \\
\hline\hline
\end{tabular}
\end{table}

\subsection{Observational data}
\label{data}

In previous work, a tension between the Union2 and Union2.1 data was usually found, because the former can present a slowing down of cosmic acceleration, while the latter fails. To test this tension, we respectively consider deceleration parameter reconstruction in these two groups of data. In addition, we also consider the GRB data in the aim of extending relevant study to high redshift. Another data we use are the OHD. In the present paper, we compile the latest OHD catalog in Table \ref{tab: Hz}.

The Union2 and Union2.1 compilations are both released by the Hubble Space Telescope Supernova Cosmology Project. Usually, they are presented as tabulated distance modulus with errors. The former, Union2 consists of 557 data points \cite{amanullah2010spectra}. While the Union2.1 contains 580 dataset \cite{suzuki2012hubble}. Their redshift regions are able to span over $z<1.414$.

The other data about the luminosity distance is the GRB which carries higher redshift than the SNIa. Unfortunately, GRBs are not standard candles like the optical supernova. In Refs. \cite{amati2006ep,ghirlanda2006gamma}, however, it was found that correlation between equivalent isotropic energy and spectral properties can be used to construct distance-redshift diagrams like the SNIa. Consequently, they certainly provide a great complement to the SNIa. Due to the lack of low-redshift long GRB data, the GRB may suffer a ``circular problem". Nevertheless, several statistical methods have been proposed to alleviate this problem \cite{ghirlanda2004gamma,firmani2005new}. Similar to calibrating SNIa using Cepheid variables, GRB also can be calibrated as standard candles via a large amount of SNIa. The key idea is the distance ladder and is cosmology-independent. Hence, a wide range of applications of GRB was arose in the cosmological constraint as a complementary
probe to SNIa \cite{wang2008model,cardone2009updated,xia2012cosmography}. In the present paper, we use the calibrated 109 GRB samples over the redshift $0.03<z<8.10$ \cite{1475-7516-2010-08-020}. They were also presented as tabulated distance modulus with errors, like the SNIa.

Following previous work \cite{seikel2012reconstruction,yahya2014null,yang2015reconstructing}, we can fix $H_0=70$ km s$^{-1}$ Mpc$^{-1}$ and include the covariance matrix with systematic errors in our calculation. To obtain the dimensionless comoving luminosity distance $D(z)$, we should make a transformation from the distance modulus via Eq. \eqref{D:define}. Moreover, the theoretical initial conditions $D(z=0)=0$ and $D'(z=0)=1$ are also taken into account in the calculation \cite{seikel2012reconstruction}. To test the influence of spatial curvature, we release it to be zero and a random variable in $\Omega_k=-0.037 \pm 0.043$ from the Planck + WP result \cite{ade2014planck}, respectively.

For the $H(z)$ data, they were not direct products from a tailored telescope, but can be acquired via two ways. Moreover, these two measurement methods are both independent of the Hubble constant. One is to calculate the differential ages of galaxies
\cite{jimenez2008constraining,simon2005constraints,stern2010cosmic}, usually called cosmic chronometer (hereafter CC $H(z)$). The other is the deduction from the BAO peaks in the galaxy power spectrum
\cite{gaztanaga2009clustering,moresco2012improved} or from the BAO
peak using the Ly$\alpha$ forest of QSOs \cite{delubac2013baryon}. For convenience, we name this type of OHD as BAO $H(z)$. In the present paper, we compile the latest dataset in Table \ref{tab: Hz}. This compilation accommodates 40 data points, which includes the most recent five measurements by \citet{moresco2016A} and the catalog in Refs. \cite{meng2015utility,moresco2016A}. After the preparation of $H(z)$ data, we should normalize them to obtain the dimensionless one $h(z)=H(z)/H_0$. Obviously, the initial condition $h(z=0)=1$ should be taken into account in our calculation. Considering the error of Hubble constant, we can calculate the uncertainty of $h(z)$
\be
\sigma_h^2 = \frac{\sigma_H^2}{H_0^2} + \frac{H^2}{H_0^4} \sigma_{H_0}^2 .
\ee
To detect the influence of Hubble constant on the reconstruction, we utilize the recent measurement $H_0=69.6 \pm 0.7$ km s$^{-1}$Mpc$^{-1}$ with 1\% uncertainty \cite{bennett20141} and $H_0=73.24 \pm 1.74$ km s$^{-1}$Mpc$^{-1}$ with 2.4\% uncertainty \cite{riess20162}, respectively.

\section{Result}
\label{result}

\begin{figure*}
\includegraphics[width=4.4cm,height=4.3cm]{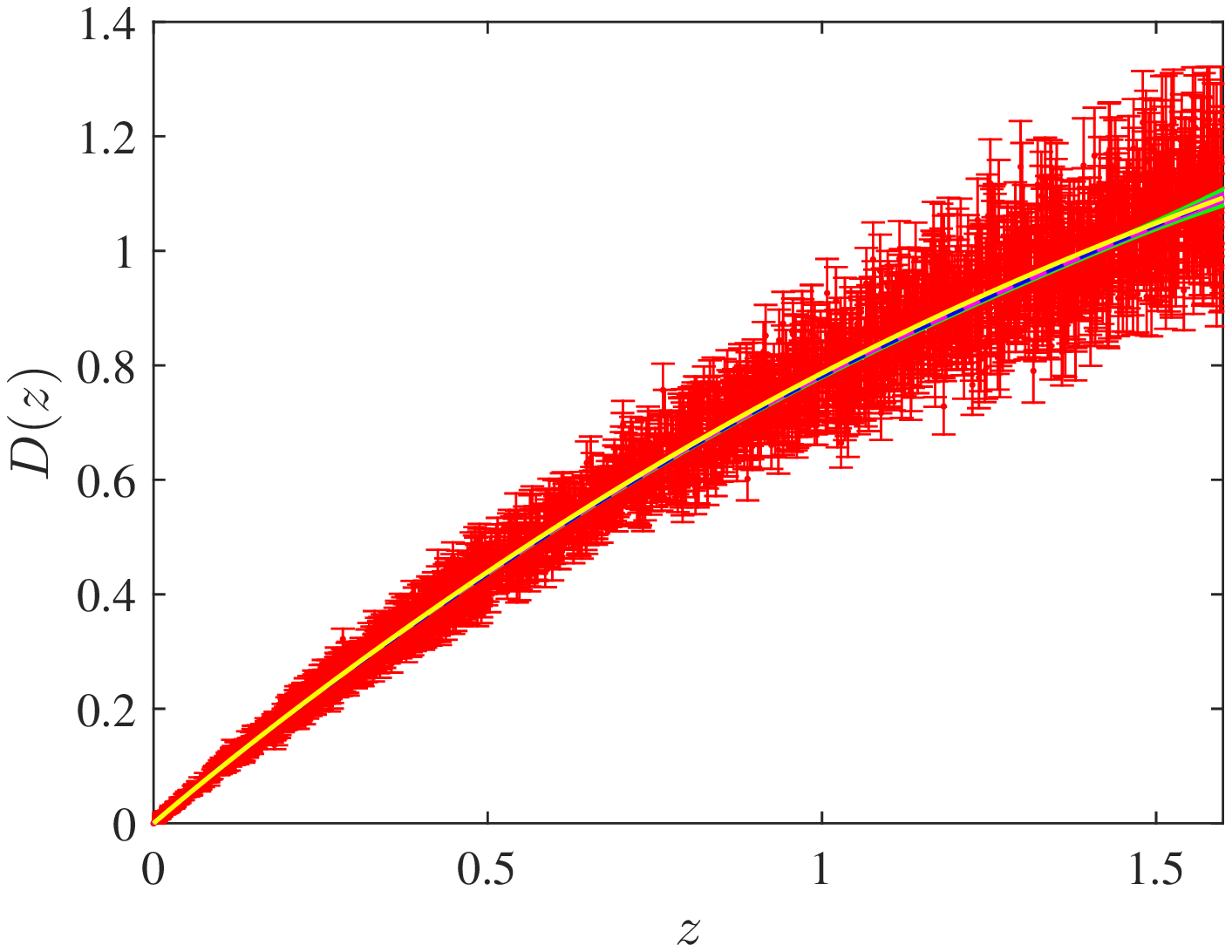}
\includegraphics[width=4.4cm,height=4.3cm]{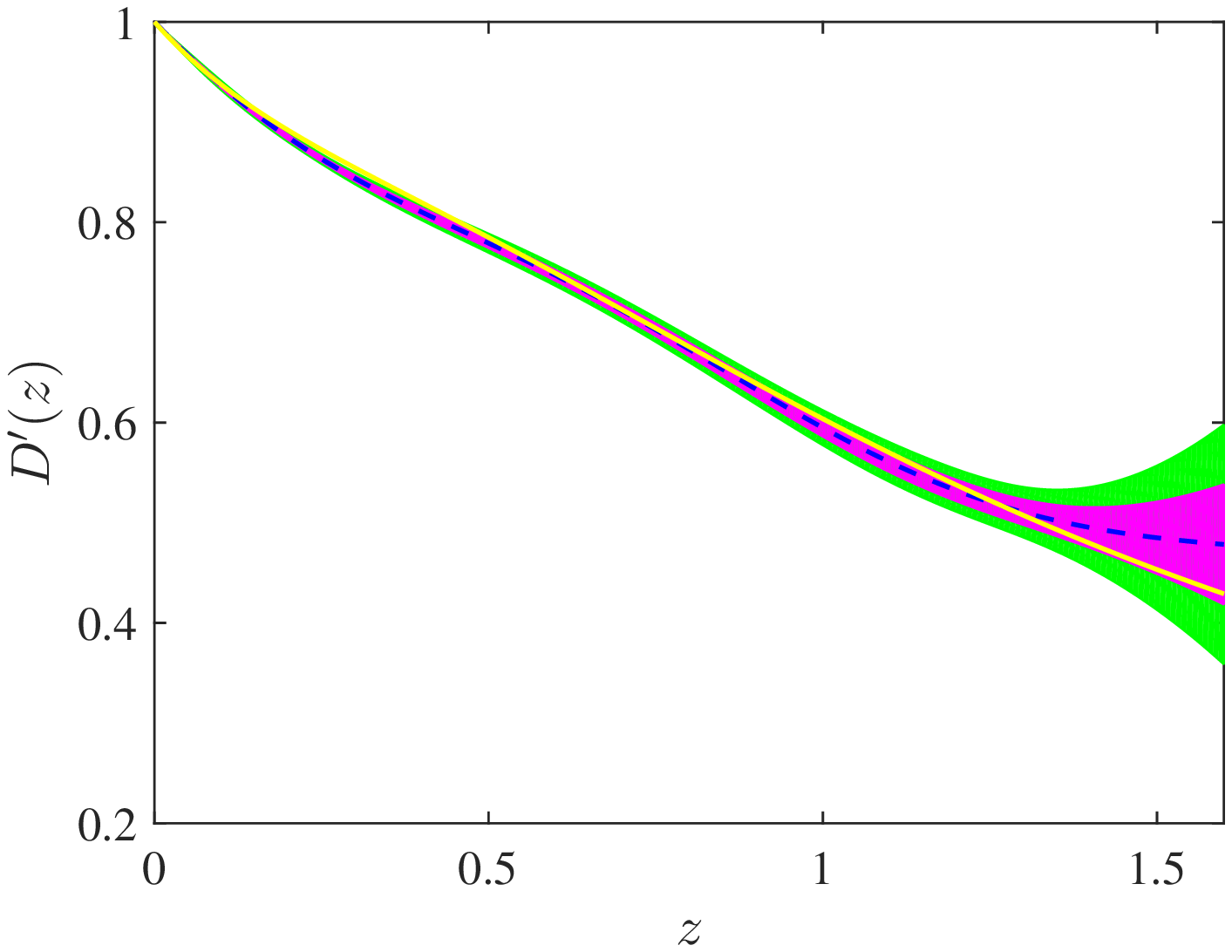}
\includegraphics[width=4.4cm,height=4.3cm]{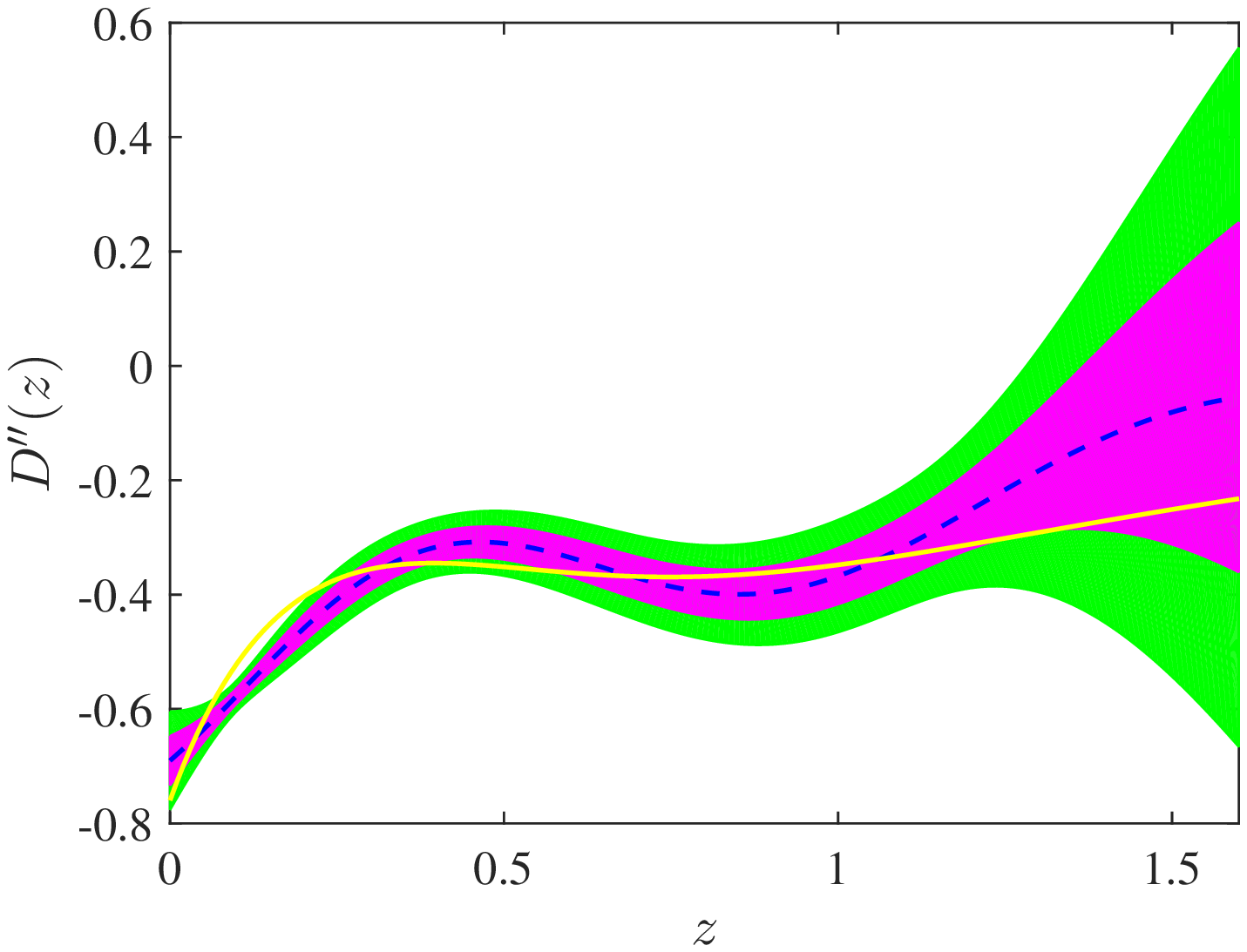}
\includegraphics[width=4.4cm,height=4.3cm]{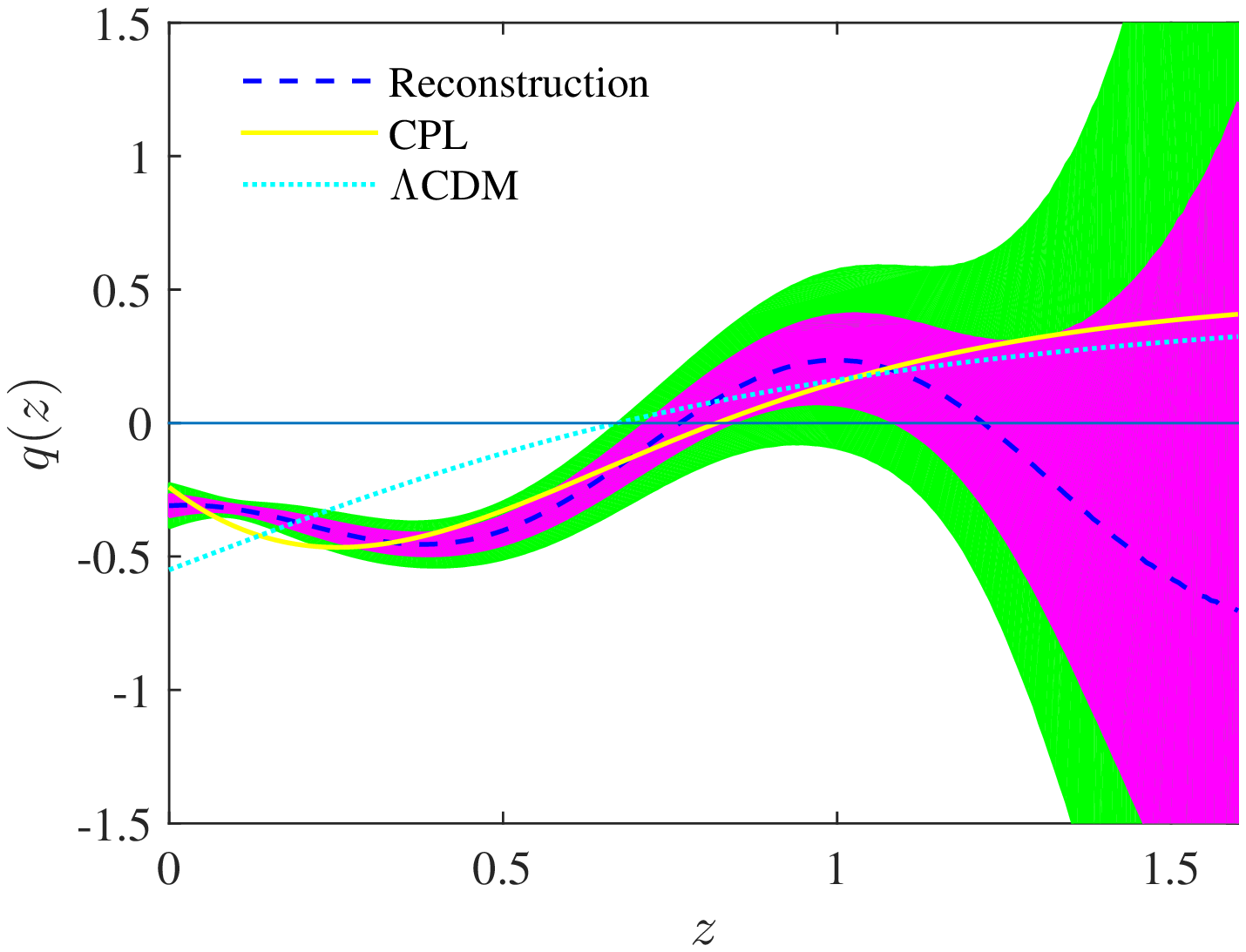}
\caption{Test of GP from the simulated supernova in the CPL model. The shaded regions are reconstructions with 68\% and 95\% confidence level. The dashed and solid lines are mean values of reconstruction and fiducial values in $\Lambda$CDM model, respectively. We compare the $\Lambda$CDM and CPL model in the rightmost panel.}  \label{test1}
\end{figure*}

\begin{figure*}
\includegraphics[width=4.4cm,height=4.3cm]{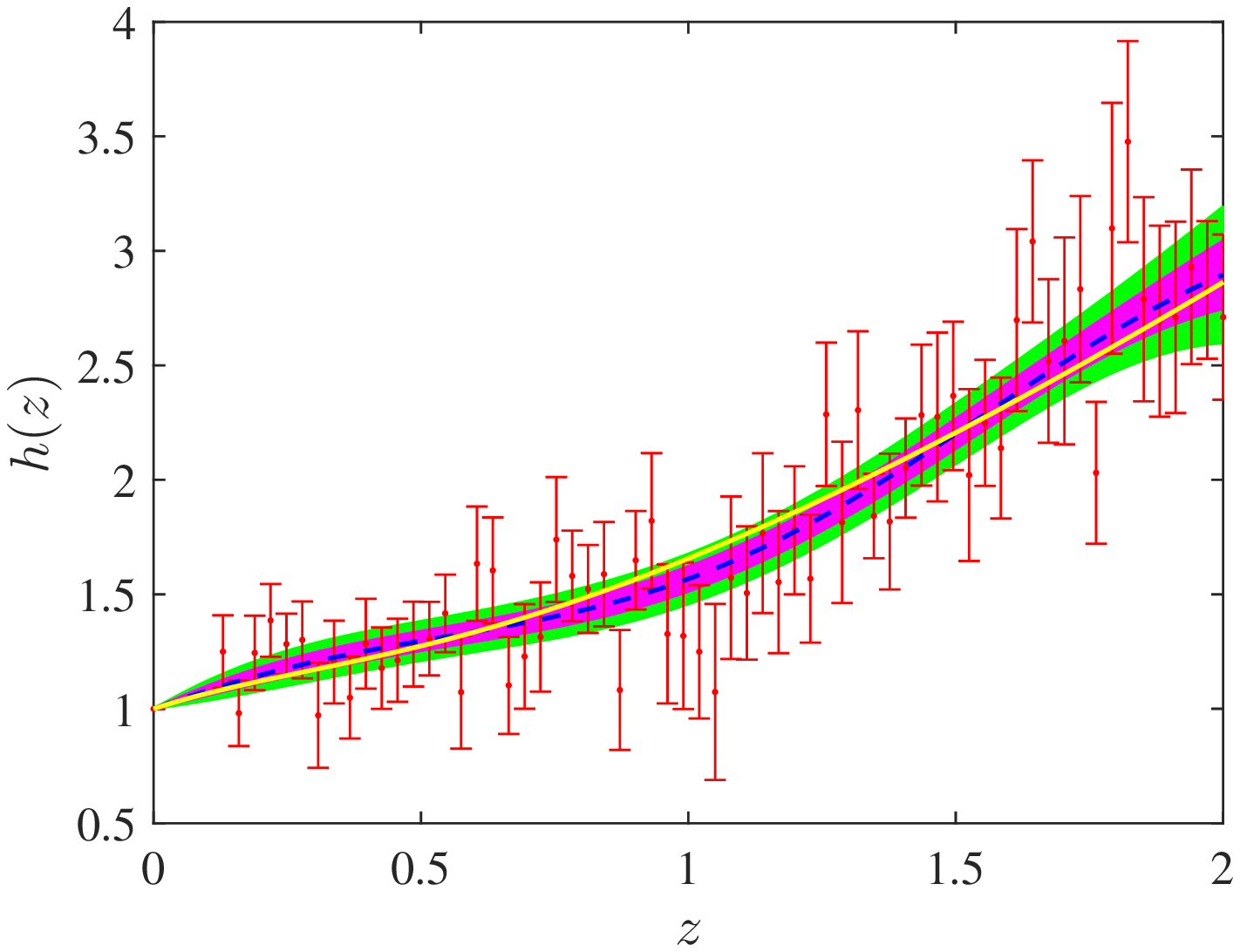}
\includegraphics[width=4.4cm,height=4.3cm]{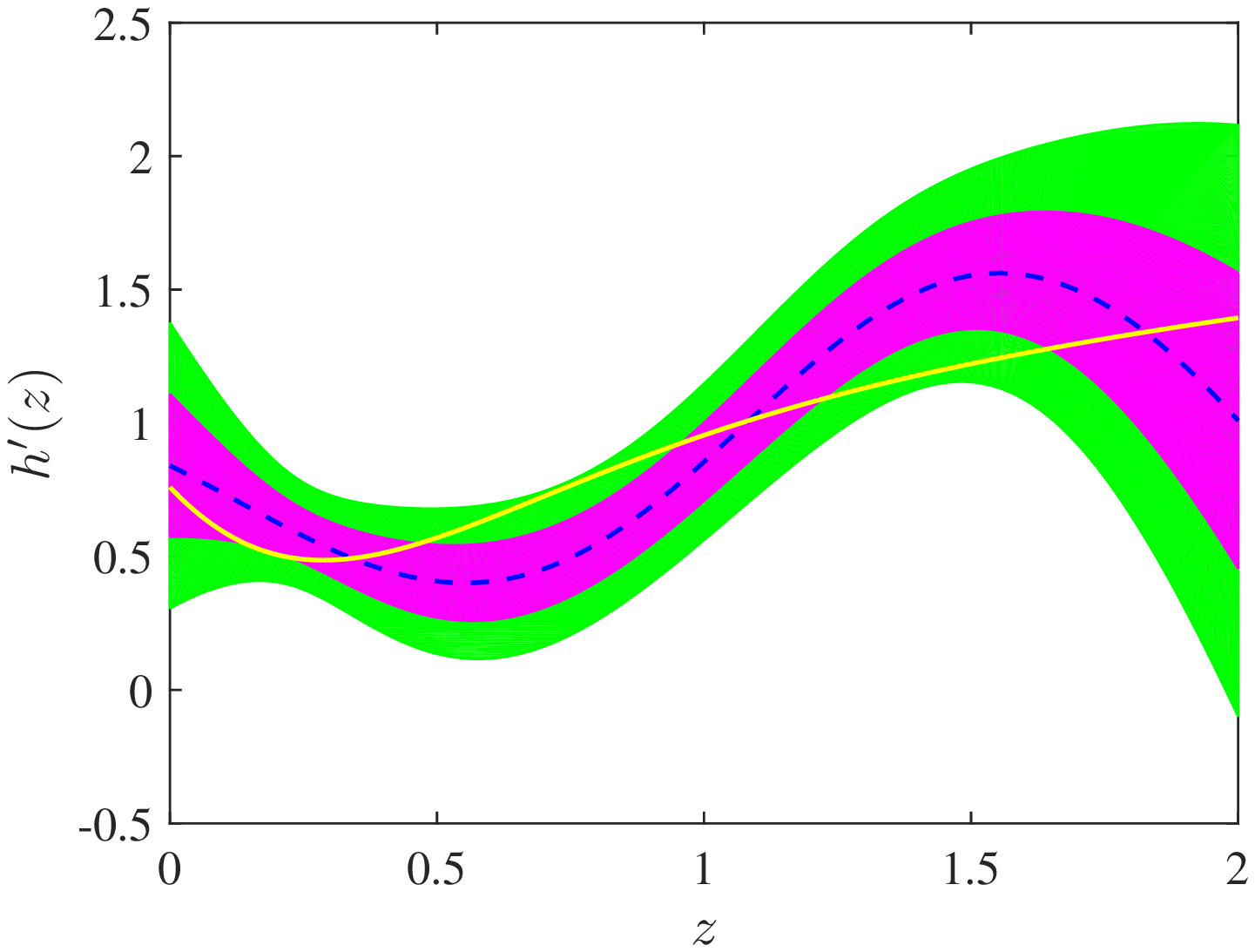}
\includegraphics[width=4.4cm,height=4.3cm]{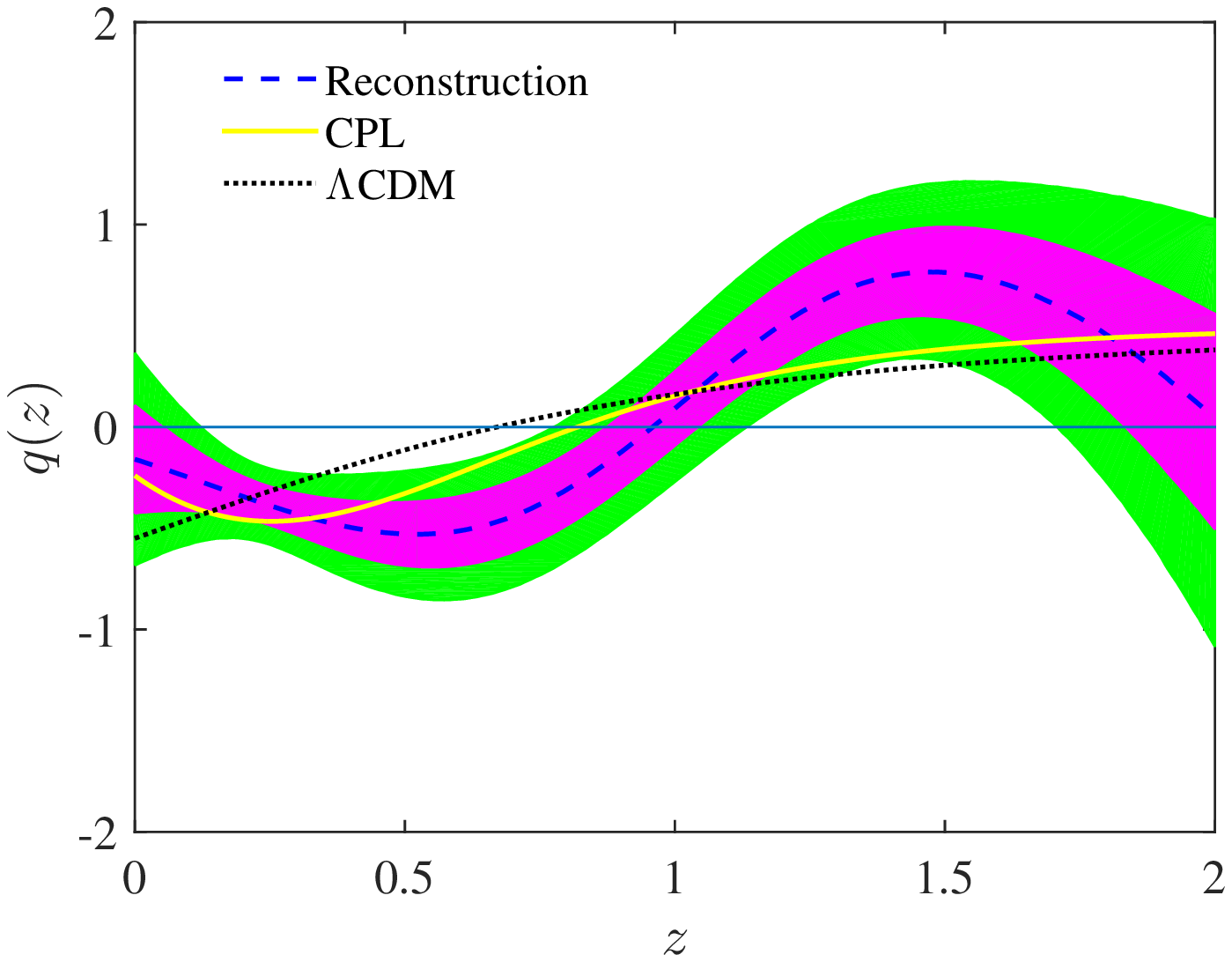}
\caption{Test of GP from the simulated $H(z)$ data in the CPL model.}  \label{test2}
\end{figure*}

\begin{figure}
    \begin{center}
\includegraphics[width=4.2cm,height=4.2cm]{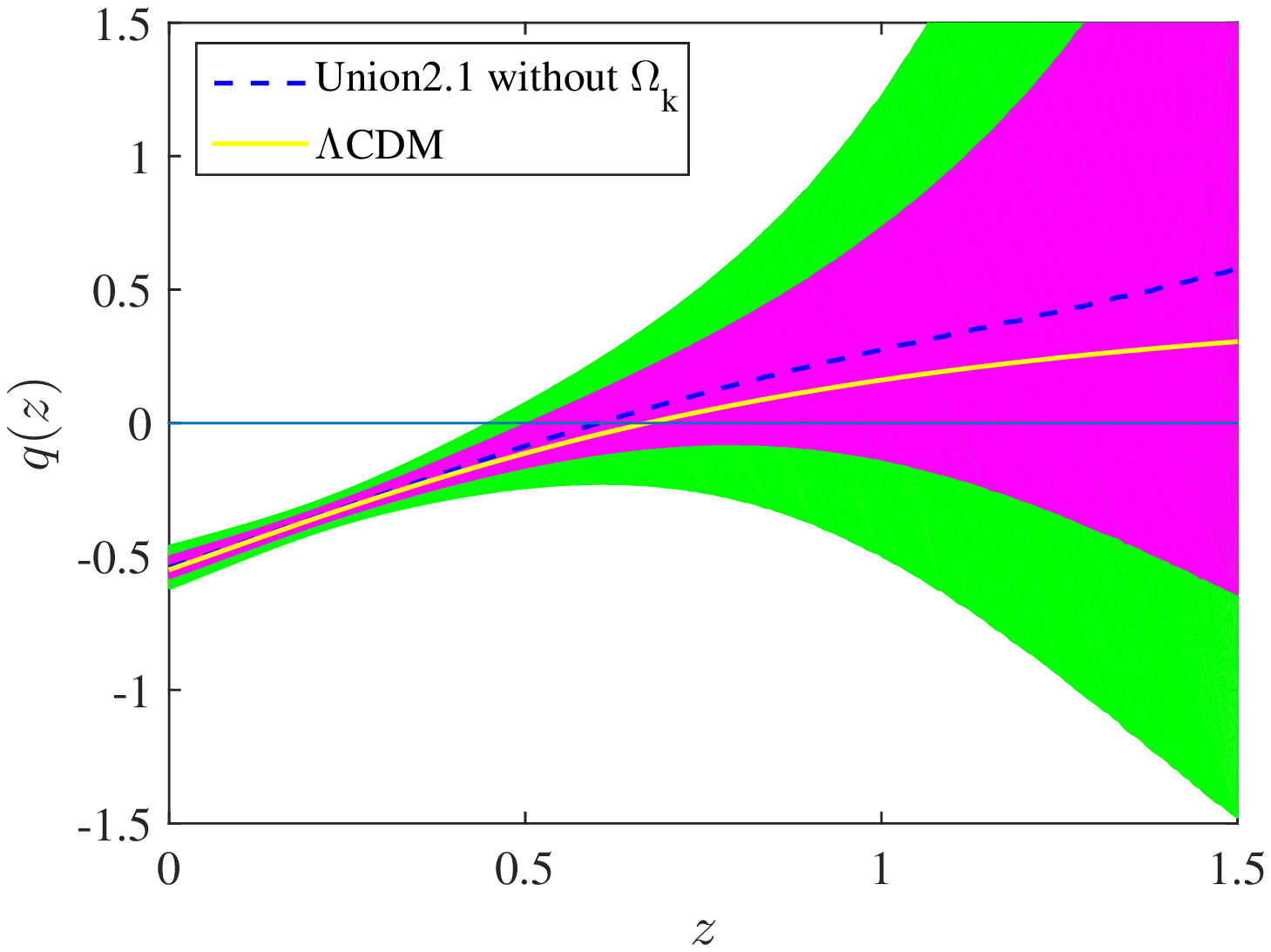}
\includegraphics[width=4.2cm,height=4.2cm]{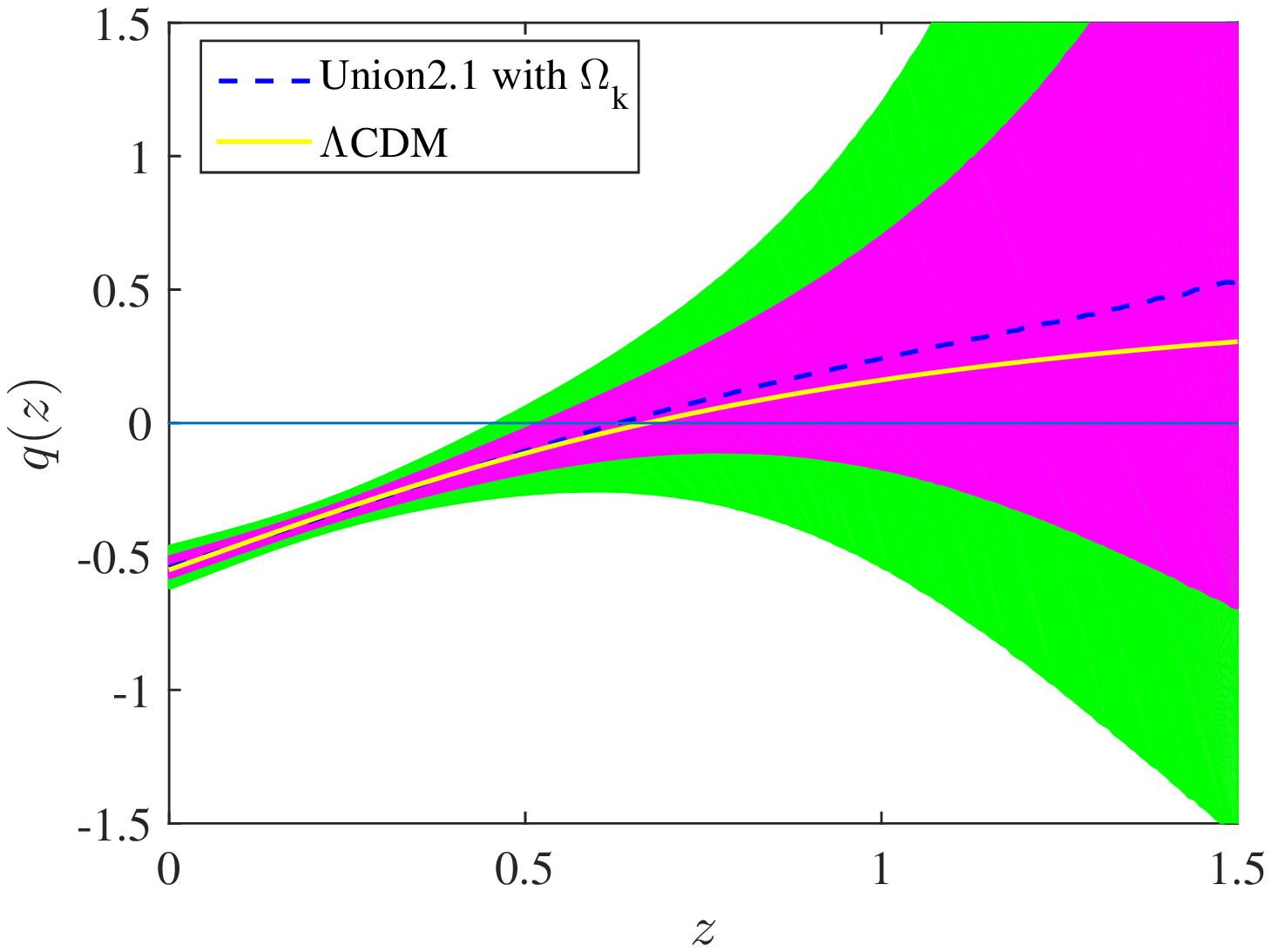}
    \end{center}
    \caption{GP reconstruction of $q(z)$ with and without spatial curvature for the Union2.1 samples. }\label{qunion21}
\end{figure}

\begin{figure}
    \begin{center}
\includegraphics[width=4.2cm,height=4.2cm]{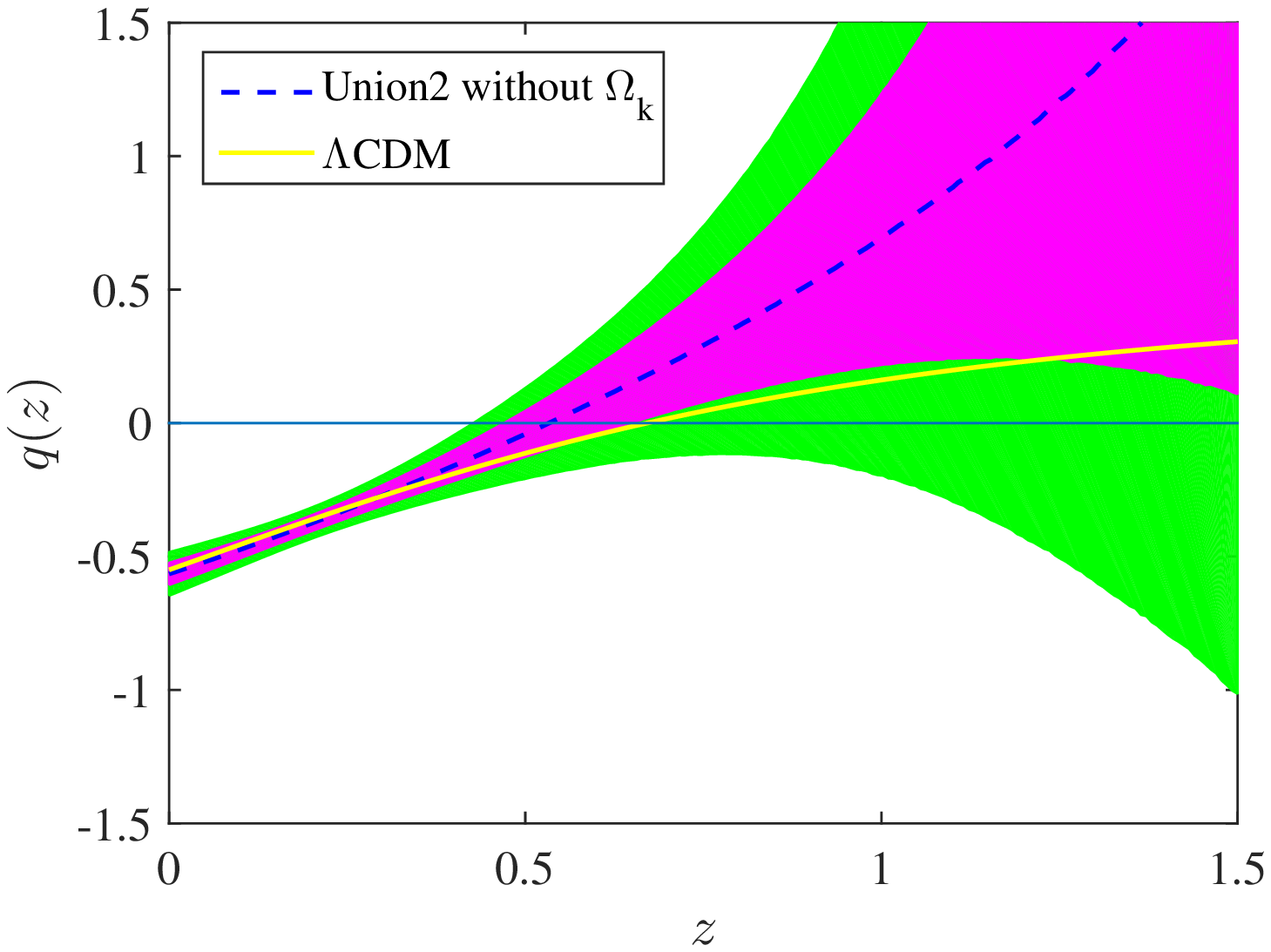}
\includegraphics[width=4.2cm,height=4.2cm]{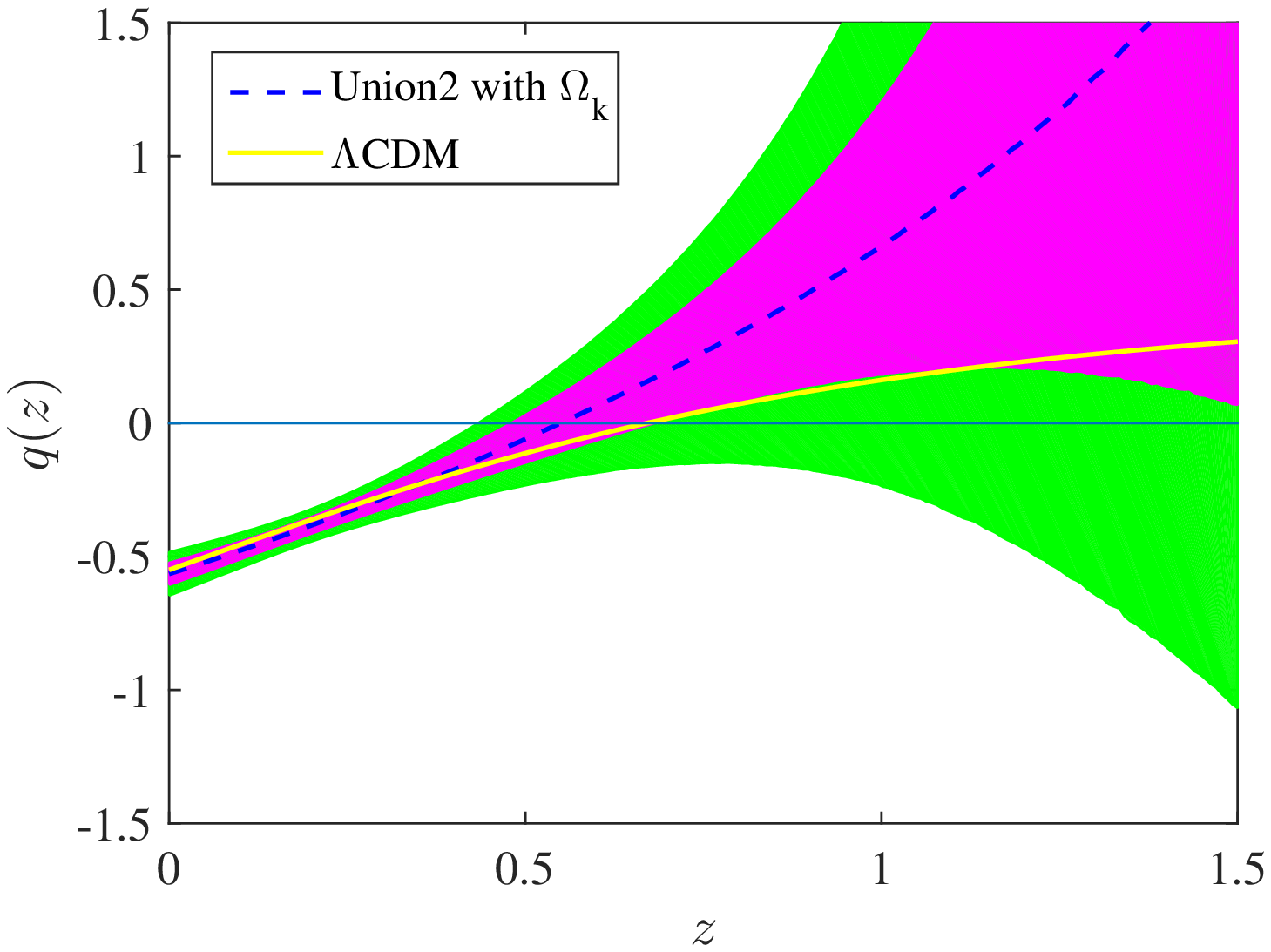}
    \end{center}
    \caption{Same as Fig. \ref{qunion21} but for the Union2 data.} \label{qunion2}
\end{figure}

\begin{figure}
    \begin{center}
\includegraphics[width=4.2cm,height=4.2cm]{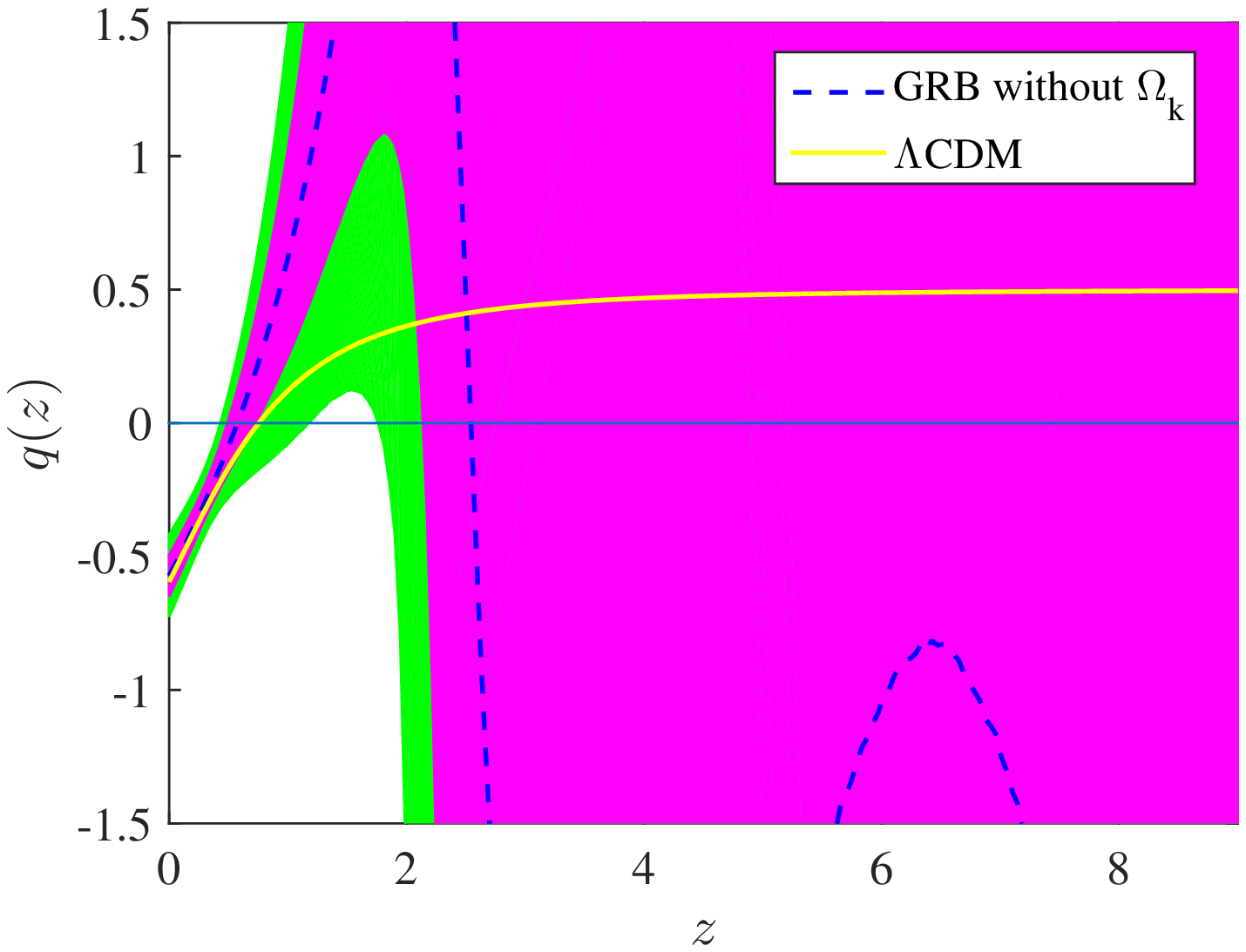}
\includegraphics[width=4.2cm,height=4.2cm]{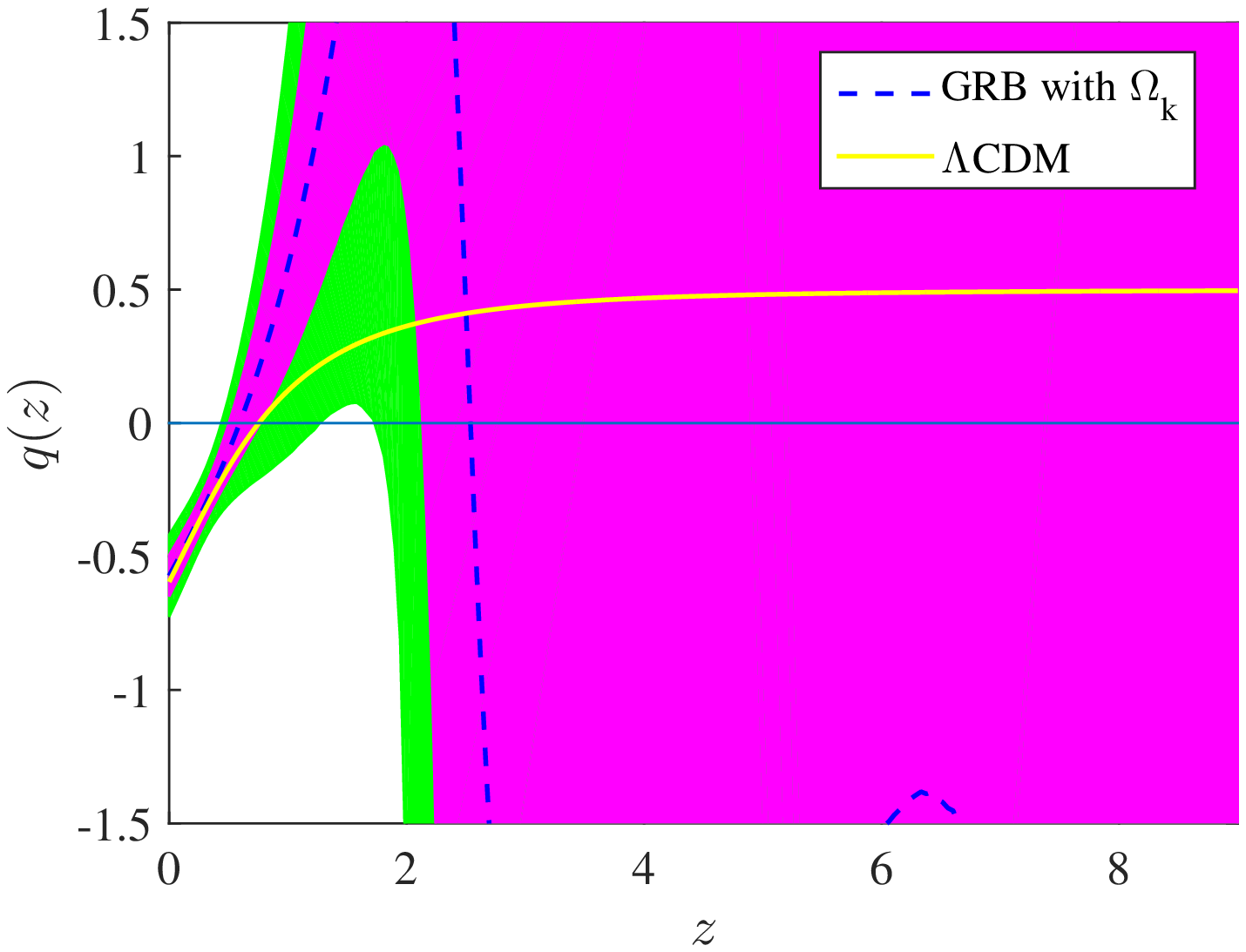}
    \end{center}
    \caption{Same as Fig. \ref{qunion21} but for the GRB data.} \label{qGRB}
\end{figure}

\subsection{Test from mock data}

Task in the section is to test the reliability of GP technique by the simulated data. If the reconstruction agrees with the fiducial value and can be distinguished from the other fiducial model, we can say that this method is useful to provide a reliable result. We show the tests from simulated SNIa in a flat universe and $H(z)$ data with one prior $H_0=73.24 \pm 1.74$ km s$^{-1}$ Mpc$^{-1}$ in Figs. \ref{test1} and \ref{test2}.

For the SNIa, Figs. \ref{test1} shows that GP reconstructions within 68\% C.L. are consistent with the fiducial CPL model, and can be distinguished from the fiducial $\Lambda$CDM model. It indicates that GP method is effective to reconstruct the deceleration parameter. We also note that errors of the reconstruction generally become bigger, especially for redshift $z>1$. It is consistent with the simulation in previous work \cite{seikel2012reconstruction,yang2015reconstructing}. This is because the lack of high-redshift data.

For the $H(z)$ data, we supply the test in Fig. \ref{test2}. We note that it does not work well like the simulated supernova data. In fact, it is not difficult to understand. First, fiducial values in the $\Lambda$CDM model and CPL model are very small, which is difficult to distinguish them, as shown in Fig. \ref{example}. Second, $H(z)$ data currently still cannot be measured by a custom-made telescope. Therefore, simulation of them only can be realized via an extrapolation method estimated from current distribution \cite{ma2011power}. Random distribution increases the difficulty of GP reconstruction obeying the fiducial CPL model. However, we find that reconstructions are all in an agreement with the fiducial values at 95\% C.L. Moreover, it is able to distinguish from the theoretical  $\Lambda$CDM model.

\subsection{Reconstruction from the Union2.1 data}
\label{resultUnion21}

GP reconstructions for the Union2.1 SNIa data are shown in Figs. \ref{qunion21}. The dashed and solid lines correspond to the mean values of reconstruction and fiducial values in the $\Lambda$CDM model, respectively.

To test the impact of spatial curvature $\Omega_k$, we plot the $q(z)$ reconstruction in two panels. Comparison shows that the spatial curvature produces slight influence on the reconstruction. First, mean value of reconstructions both agree well with the fiducial model, but a small variation in the high redshift. Second, difference of transition redshift from deceleration expansion to acceleration for different spatial curvature is slight, as shown in Table \ref{tab: transition}. In addition, we cannot get the upper bound of the transition redshift. This may be caused by the large error of observational data at high redshift. Last, but the most importantly, slowing down of cosmic acceleration within 95\% C.L. for any spatial curvature cannot be favored by the Union2.1 supernova data. Also, no peak of the cosmic acceleration is found. Certainly, this estimation is in a full agreement with previous work. In addition, the GP reconstruction gives a current deceleration parameter $q_0=-0.54$ as well as the fiducial value.

\subsection{Reconstruction from the Union2 data}
\label{resultUnion21}

Fig. \ref{qunion2} plots the $q(z)$ reconstruction for Union2 data with different spatial curvature. In previous work, some of them found that the slowing down of cosmic acceleration may be supported by the Union2 supernova data. Our reconstruction shows that no convincing evidence can be offered to favor this speculation. Moreover, the spatial curvature imposes no significant influence on the reconstruction. Transition redshift in these two cases are also fairly consistent, as shown in Table \ref{tab: transition}. In following section, we plan to explore the reason of tension between GP reconstruction and dark energy parametrization.

\begin{table}
\caption{\label{tab: transition} Transition redshift for different data from deceleration to acceleration. The $H_0$ is in the unit of km s$^{-1}$Mpc$^{-1}$.  }
\begin{tabular}{ccc}
\hline
data   & $\Omega_k = 0$ &  $\Omega_k \neq 0$   \\
\hline
Union2.1      & $z_t = 0.60^{+null}_{-0.09}$    & $z_t = 0.52^{+null}_{-0.11}$   \\
Union2        & $z_t = 0.53^{+0.11}_{-0.06}$    & $z_t = 0.55^{+0.13}_{-0.07}$    \\
GRB           & $z_t = 0.58^{+0.15}_{-0.08}$    & $z_t = 0.59^{+0.17}_{-0.09}$    \\
\hline
\hline
data       &$H_0=69.60\pm0.07$  &$H_0=73.24\pm1.74$      \\
\hline
CC    &$z_t = 0.56^{+0.17}_{-0.10}$    & $z_t = 0.54^{+0.12}_{-0.08}$     \\
BAO   &$z_t = 0.80^{+0.18}_{-0.15}$    & $z_t = 0.82^{+0.17}_{-0.15}$ \\
total & $z_t = 0.68^{+0.12}_{-0.09}$    & $z_t = 0.67^{+0.11}_{-0.08}$   \\
\hline
\end{tabular}
\end{table}

\subsection{Reconstruction from the GRB data}
\label{resultGRB}

GRB data contain less samples than Union2 and Union2.1 data, but with higher redshift. Such data may provide a complement to previous study of the slowing down of acceleration. In Fig. \ref{qGRB}, we show the $q(z)$ reconstruction for different spatial curvature. We find that GRB data present a rough result with much larger uncertainty, especially at high redshift. Impact of the spatial curvature in this case is also so trivial. For the transition redshift in Table \ref{tab: transition}, on the one hand, spatial curvature presents little impact on it. On the other hand, its error is small like the Union2 data. One is because GRB data should be calibrated via the low-redshift supernova data in a fiducial model. The other is that only statistical uncertainties were issued. Same as the supernova data, slowing down of acceleration still cannot be evidenced within 95\% C.L.


\begin{figure}
\includegraphics[width=4.2cm,height=4.3cm]{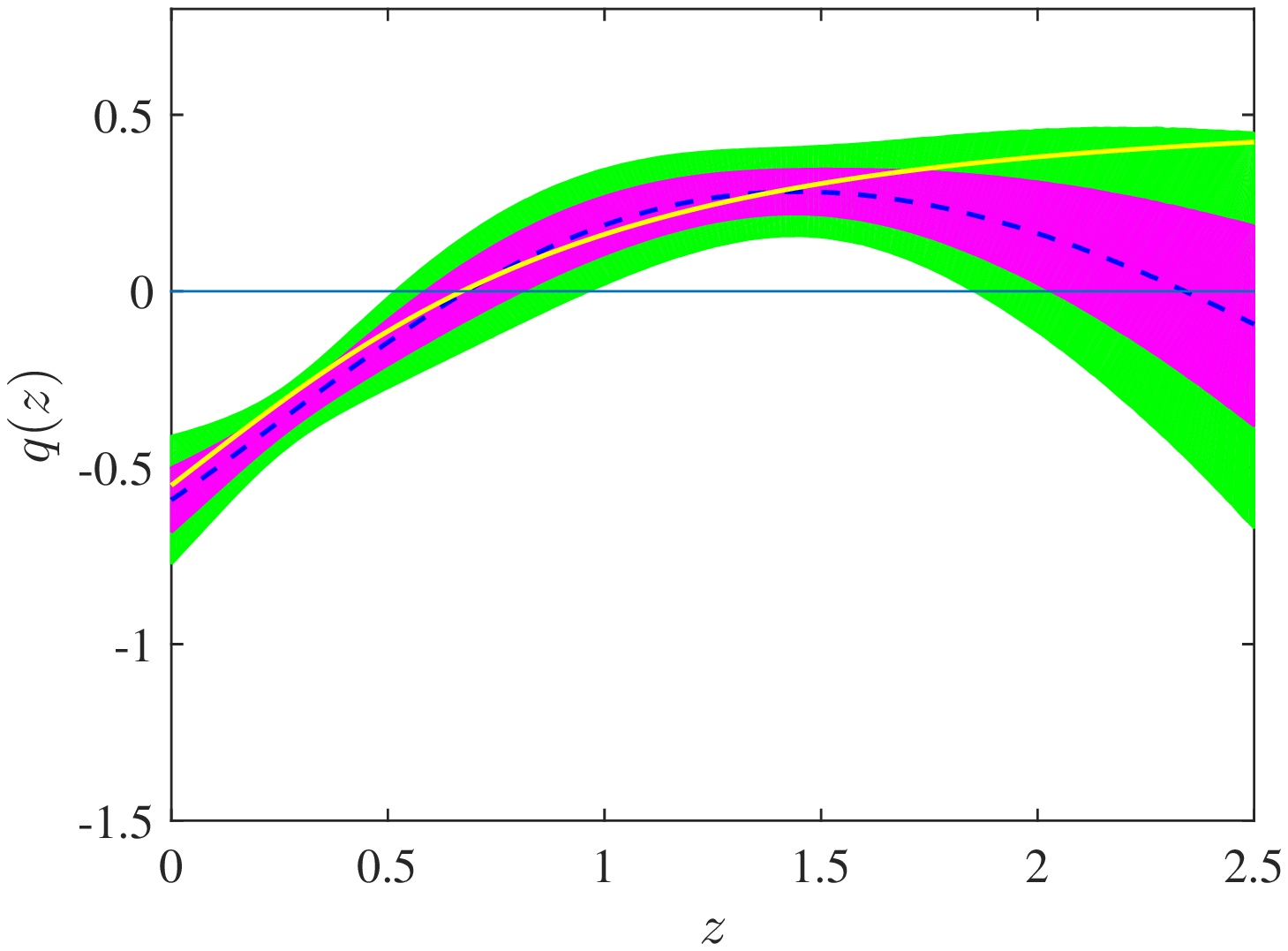}
\includegraphics[width=4.2cm,height=4.3cm]{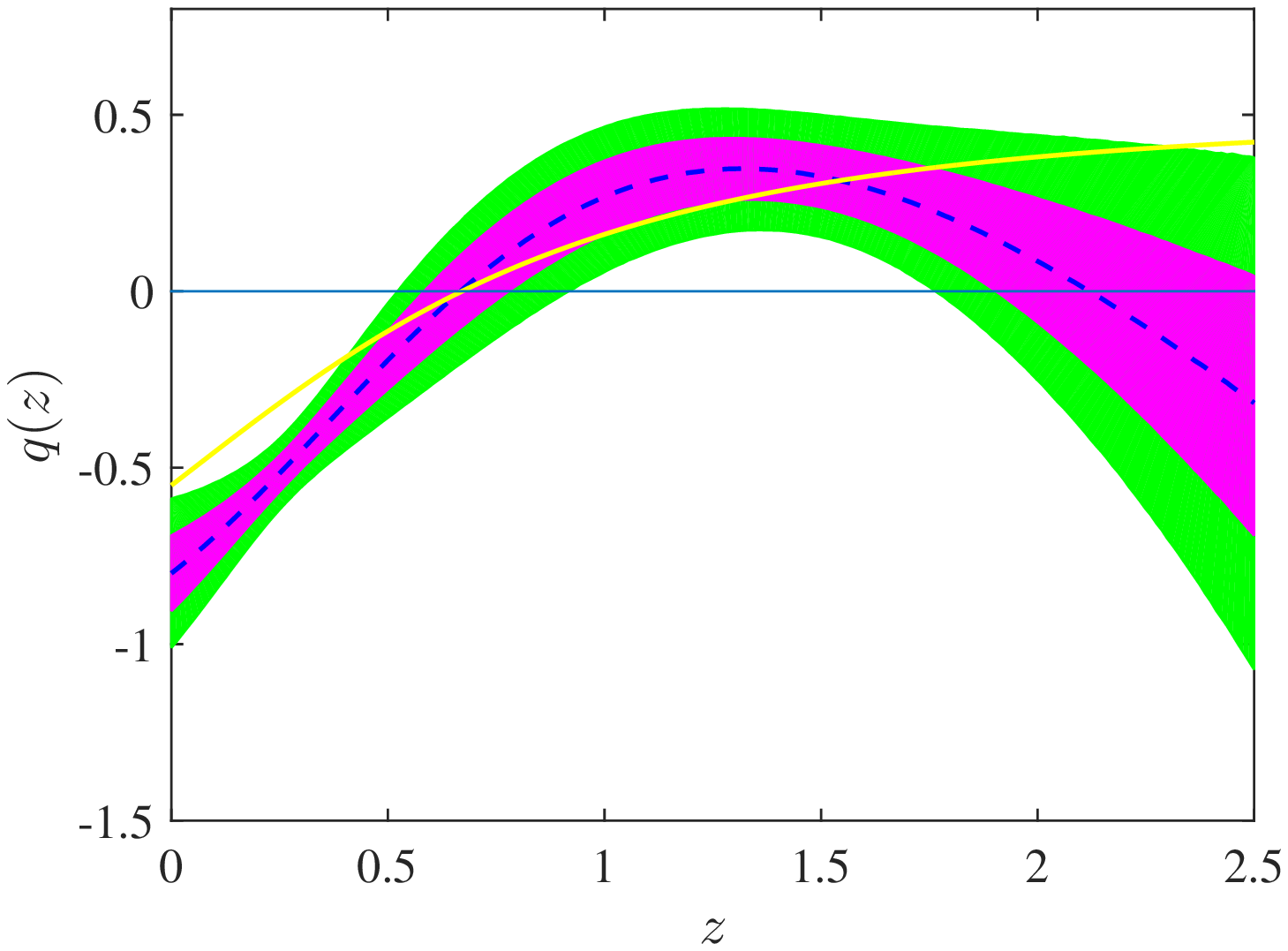}
\caption{GP reconstruction of $q(z)$ for the total $H(z)$ data with $H_0=69.60 \pm 0.7$ km s$^{-1}$ Mpc$^{-1}$ (\textit{left panel}) and $H_0=73.24 \pm 1.74$ km s$^{-1}$ Mpc$^{-1}$ (\textit{right panel}). }  \label{total_OHD1}
\end{figure}

\subsection{Reconstruction from the OHD}
\label{resultOHD}

\begin{figure}
\includegraphics[width=4.2cm,height=4.3cm]{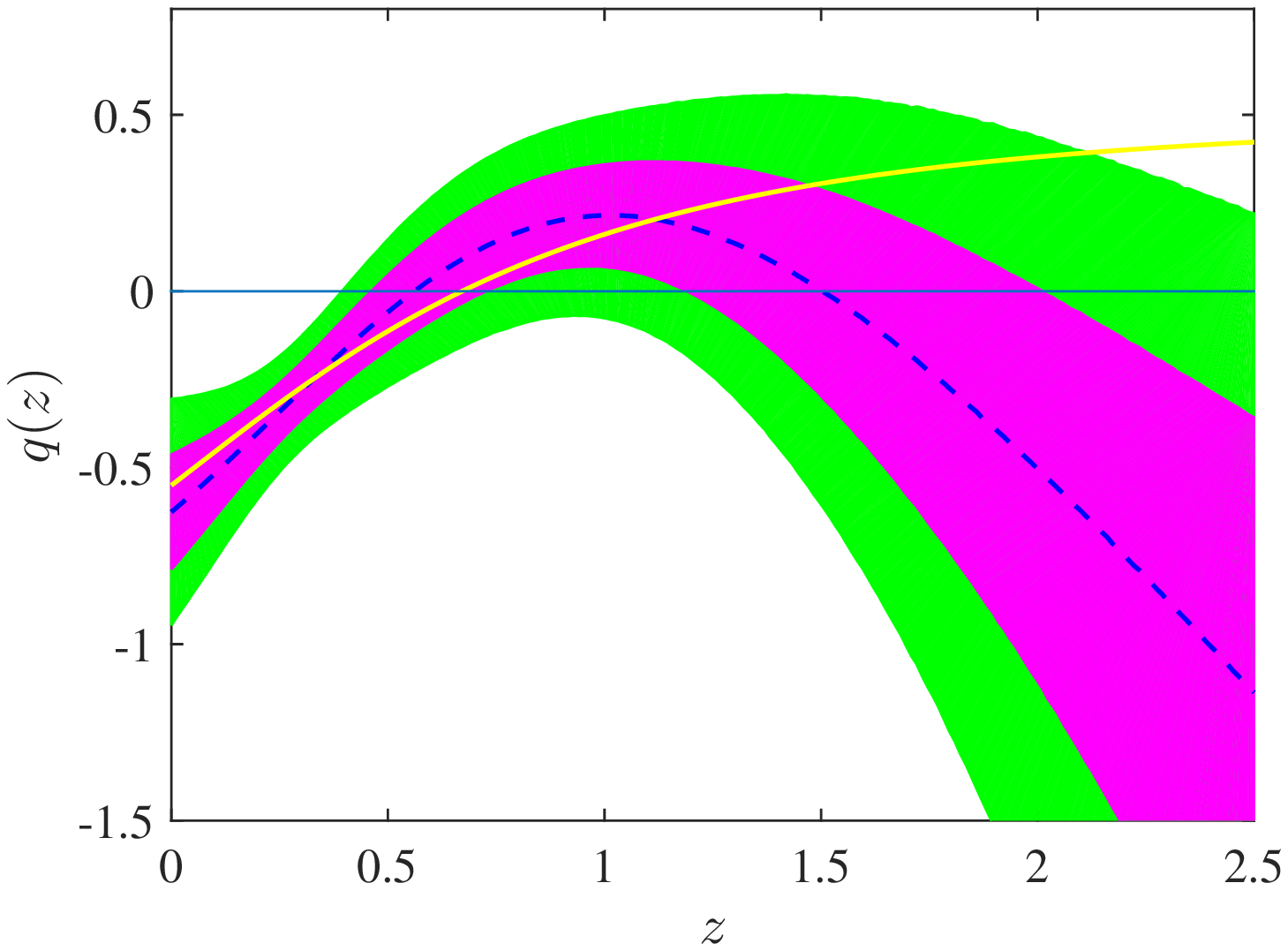}
\includegraphics[width=4.2cm,height=4.3cm]{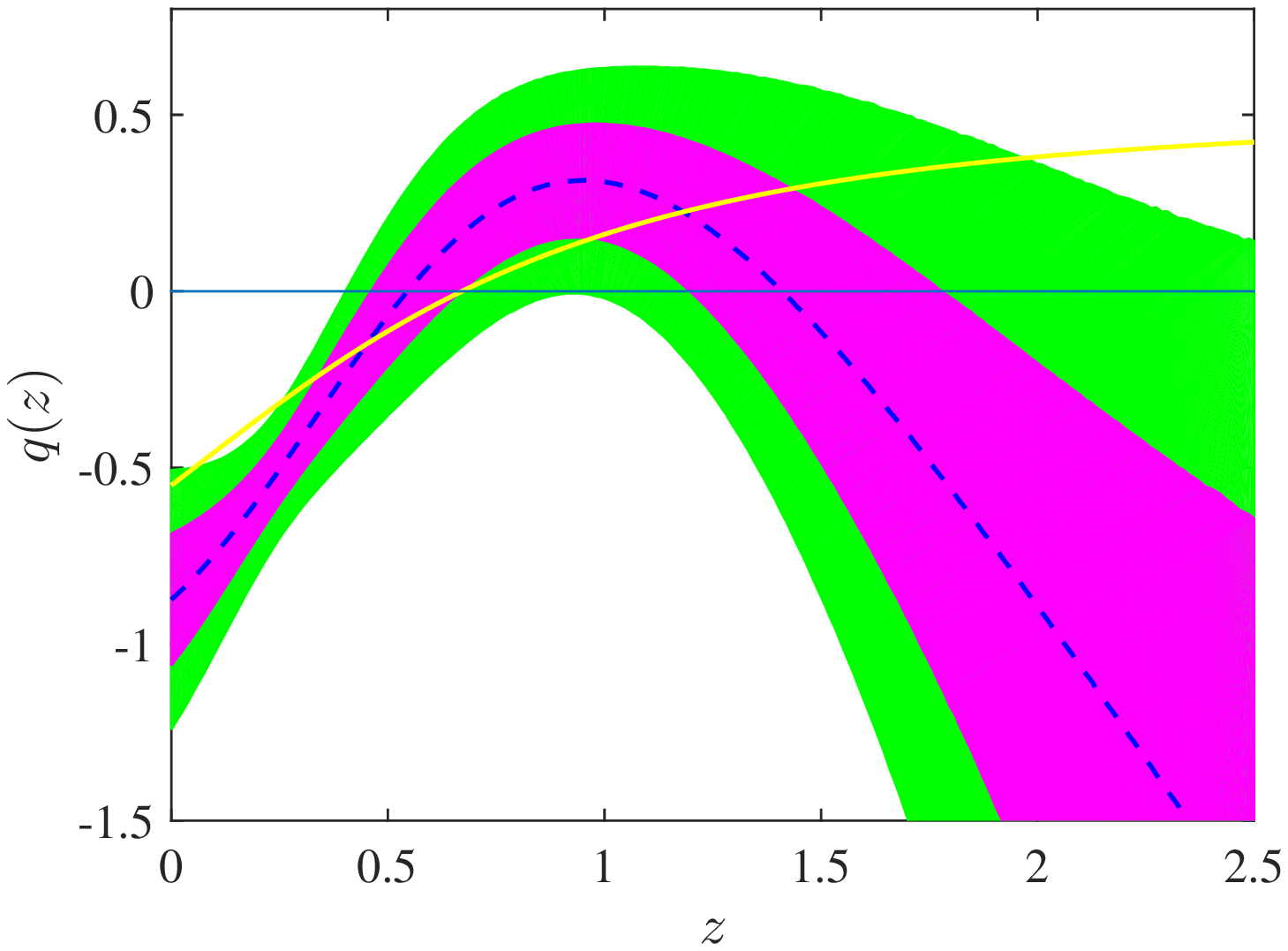}
\caption{Same as Fig. \ref{total_OHD1} but for the CC $H(z)$ data. }  \label{OHD_CC}
\end{figure}
\begin{figure}
\includegraphics[width=4.2cm,height=4.3cm]{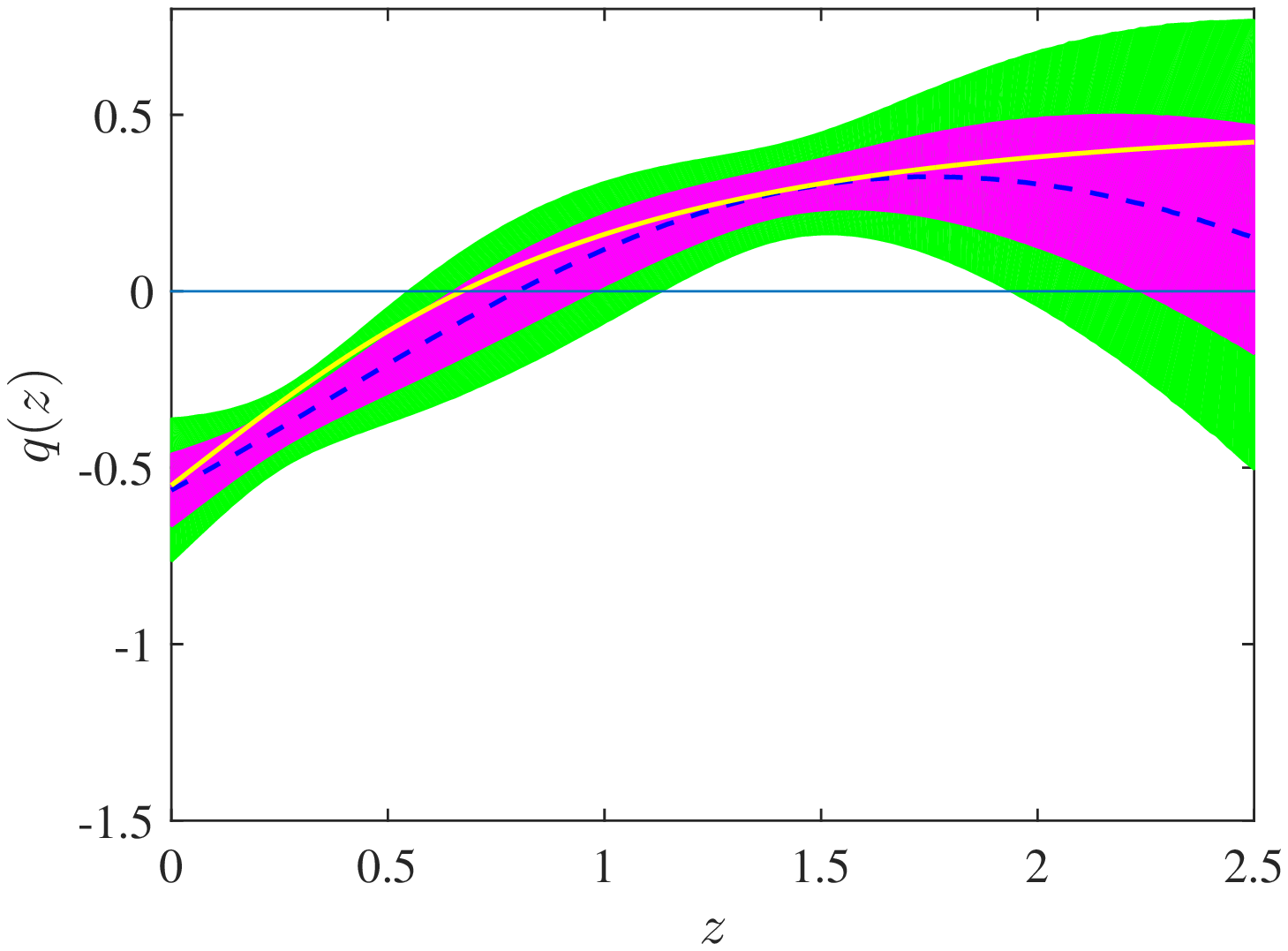}
\includegraphics[width=4.2cm,height=4.3cm]{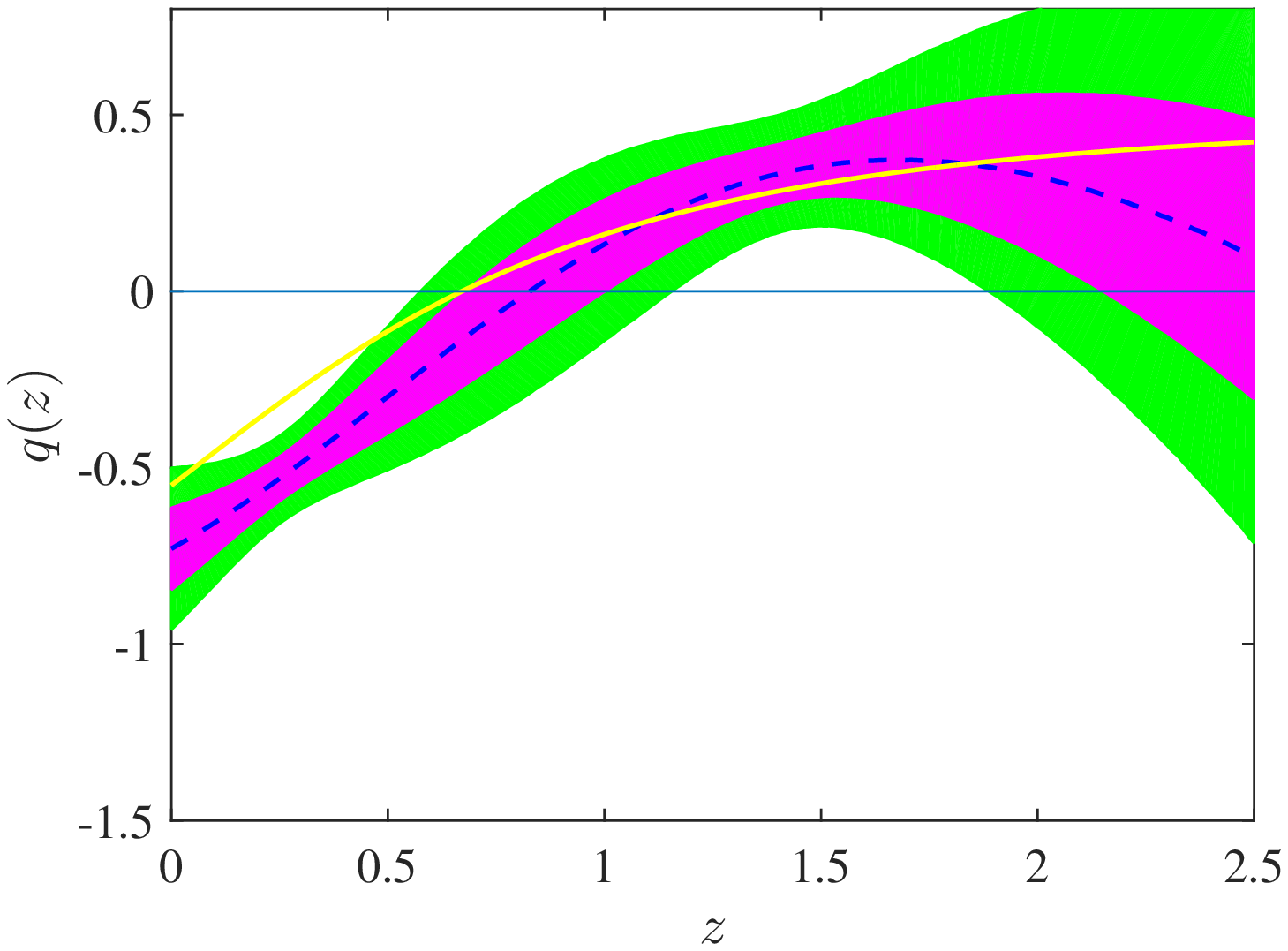}
\caption{Same as Fig. \ref{total_OHD1} but for the BAO $H(z)$ data.}  \label{OHD_BAO}
\end{figure}

In this section, we report the $q(z)$ reconstruction using OHD and take into account two priors of $H_0$. To test the influence of $H_0$, we use the fiducial $\Lambda$CDM model as a tool. Moreover, we study the potential impact of different $H(z)$ samples, i.e., CC $H(z)$ data and BAO $H(z)$ data on the results.

In Fig. \ref{total_OHD1}, we plot the $q(z)$ reconstruction for different $H_0$. Although \citet{seikel2012using} also investigated the constraint by OHD and showed them in their Figure 1, we collect more data and consider the impact of $H_0$ in the present paper, which may influence the relevant reconstruction. First, we note an obviously departures from the fiducial model for higher redshift $z>1.5$. In following, we will find that this is caused by the CC $H(z)$ data. Second, comparison shows that $H_0$ strongly affects the profile of reconstruction, especially for low redshift, but does not influence the transition redshift as shown in Table \ref{tab: transition}. Finally, no evidence hints that the cosmic acceleration has reached its peak.

$q(z)$ for the CC $H(z)$ data are shown in Fig. \ref{OHD_CC}. Evidence shows that no slowing down of cosmic acceleration is found. Table \ref{tab: transition} shows that they present a similar transition redshift as the supernova data. We also note that reconstruction at high redshift is obviously against the fiducial value. As well as the total OHD, Hubble constant in this case also influence the profile of reconstruction.

We show the reconstruction for BAO $H(z)$ data in Fig. \ref{OHD_BAO}. Seriously, slowing down of acceleration still cannot be obtained within 95\% C.L. However, a bigger transition redshift is given. Comparing with the Fig. \ref{OHD_CC}, they present a much different profile from the CC $H(z)$ data, especially for high redshift. Transition redshift in the two prior of $H_0$ are similar, but bigger than those for the CC $H(z)$ data.

\section{Tension analysis for Union2 data}
\label{Union2_discussion}

\begin{figure*}
\includegraphics[width=4.4cm,height=4.3cm]{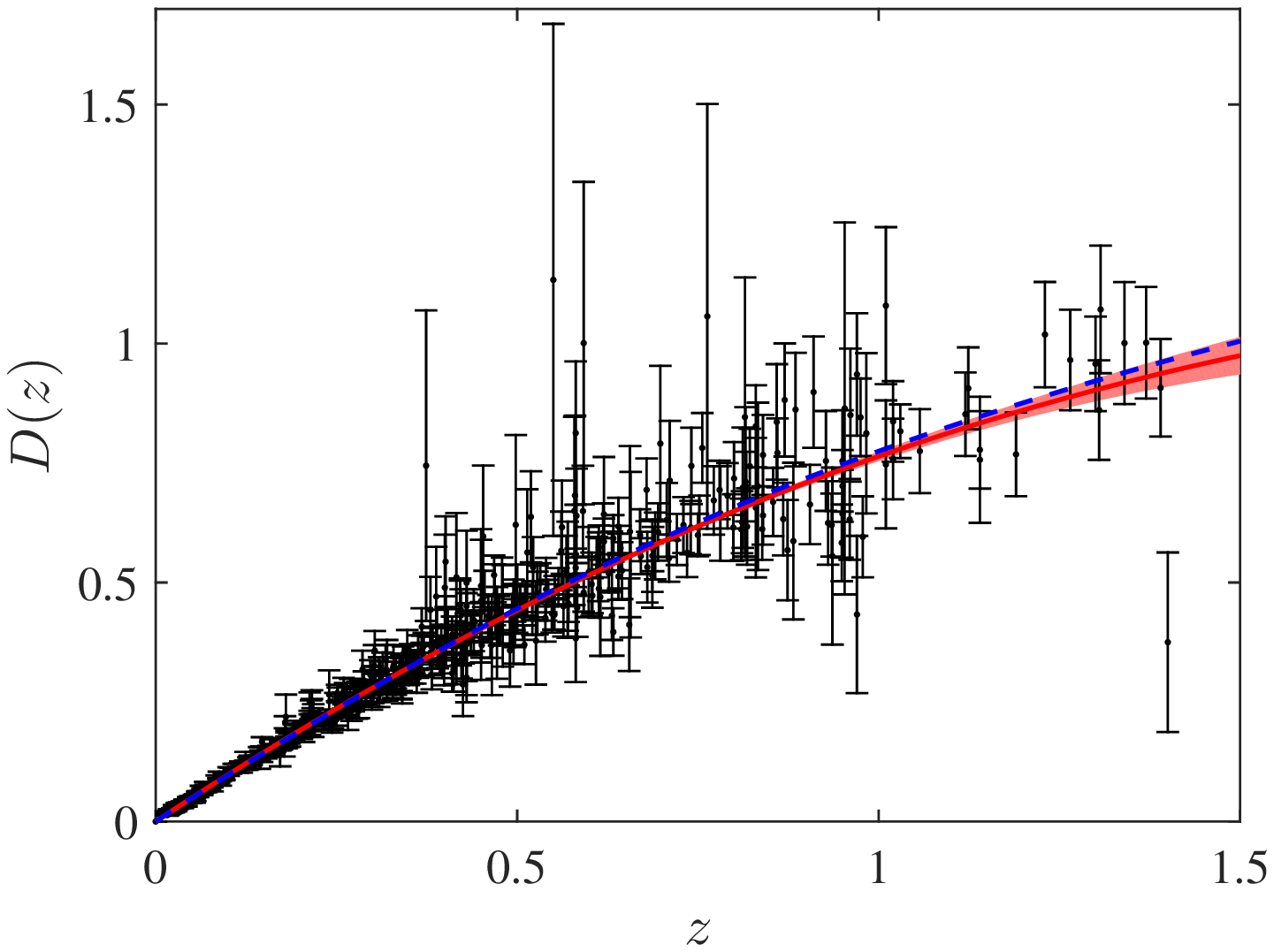}
\includegraphics[width=4.4cm,height=4.3cm]{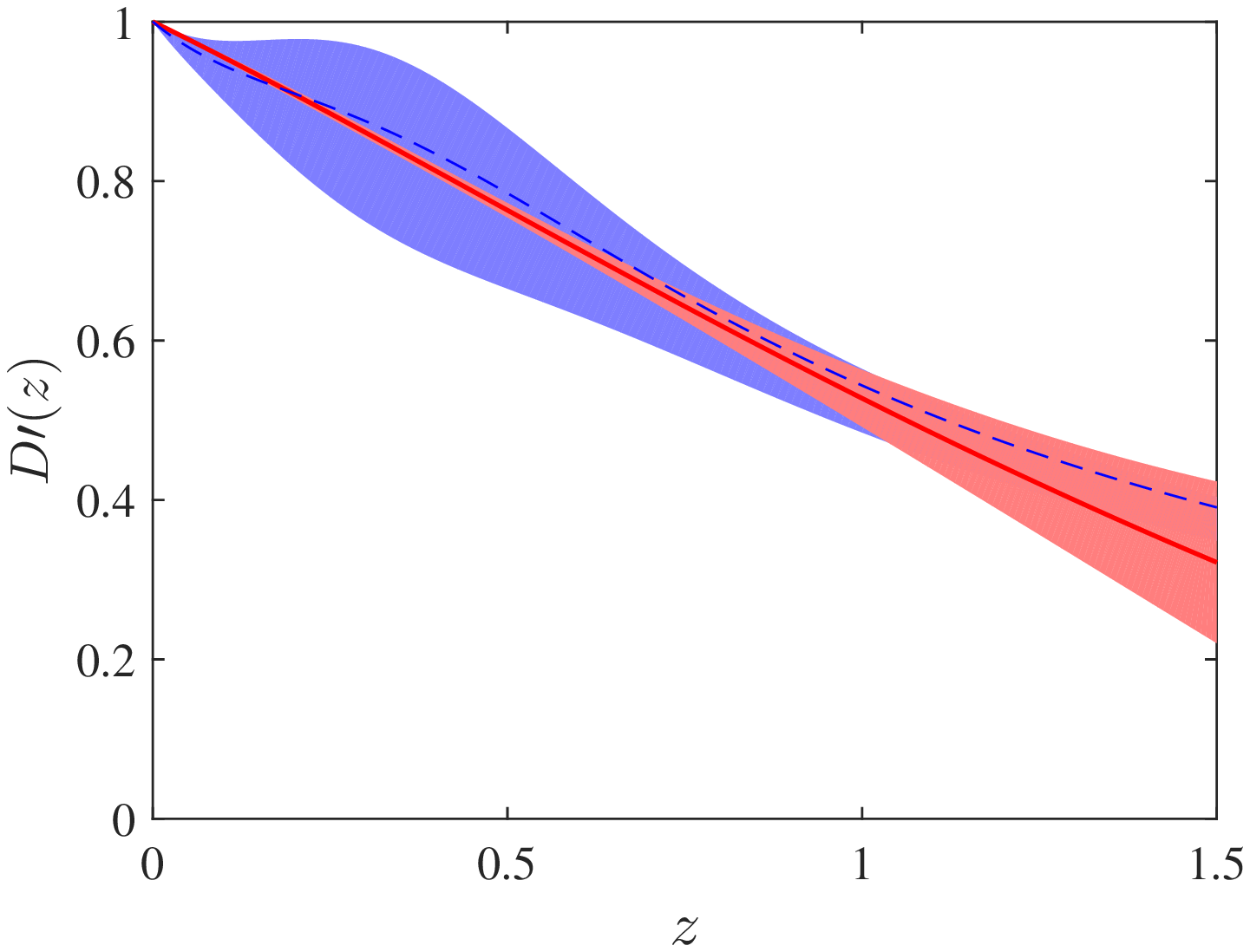}
\includegraphics[width=4.4cm,height=4.3cm]{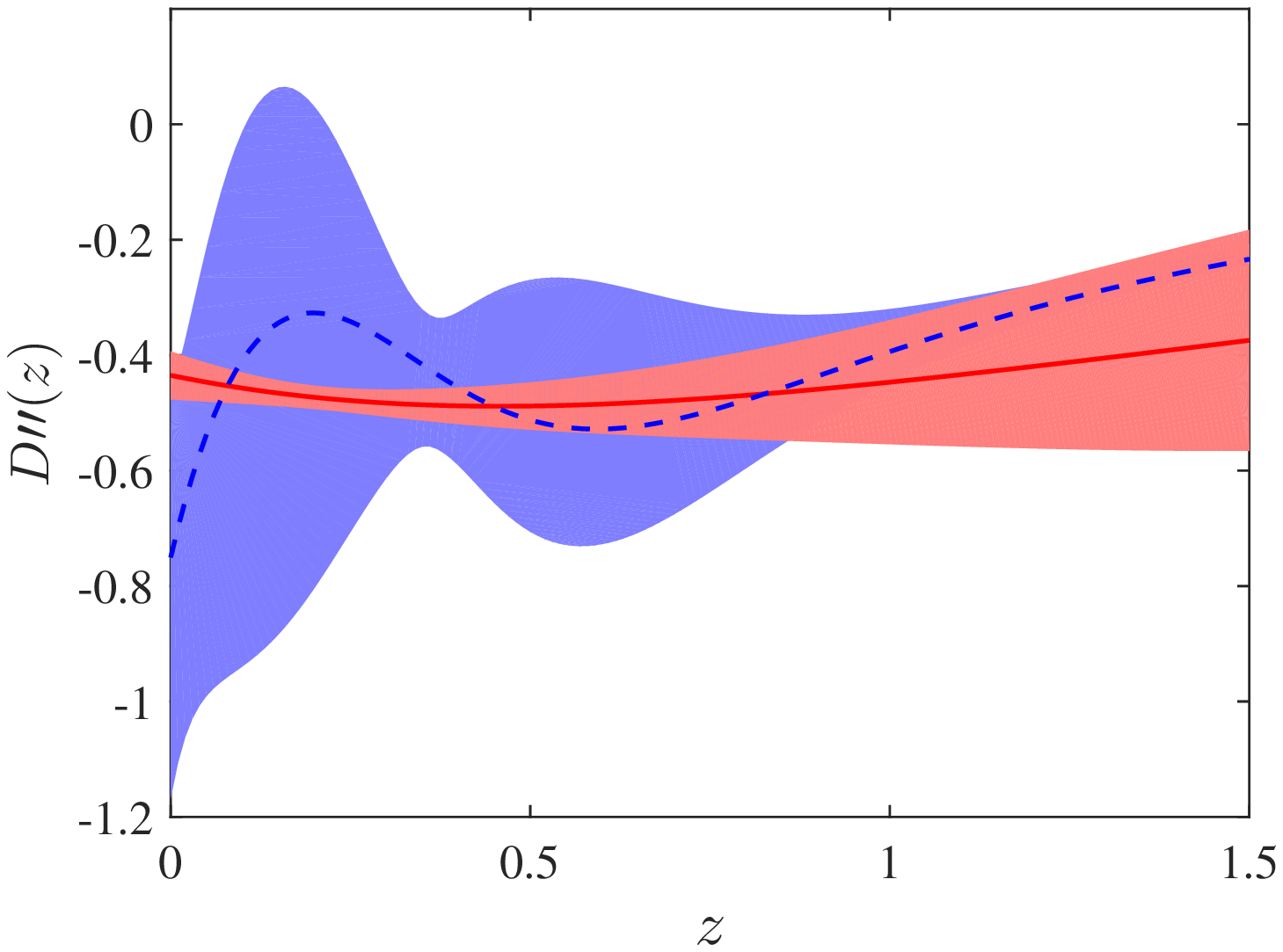}
\includegraphics[width=4.4cm,height=4.3cm]{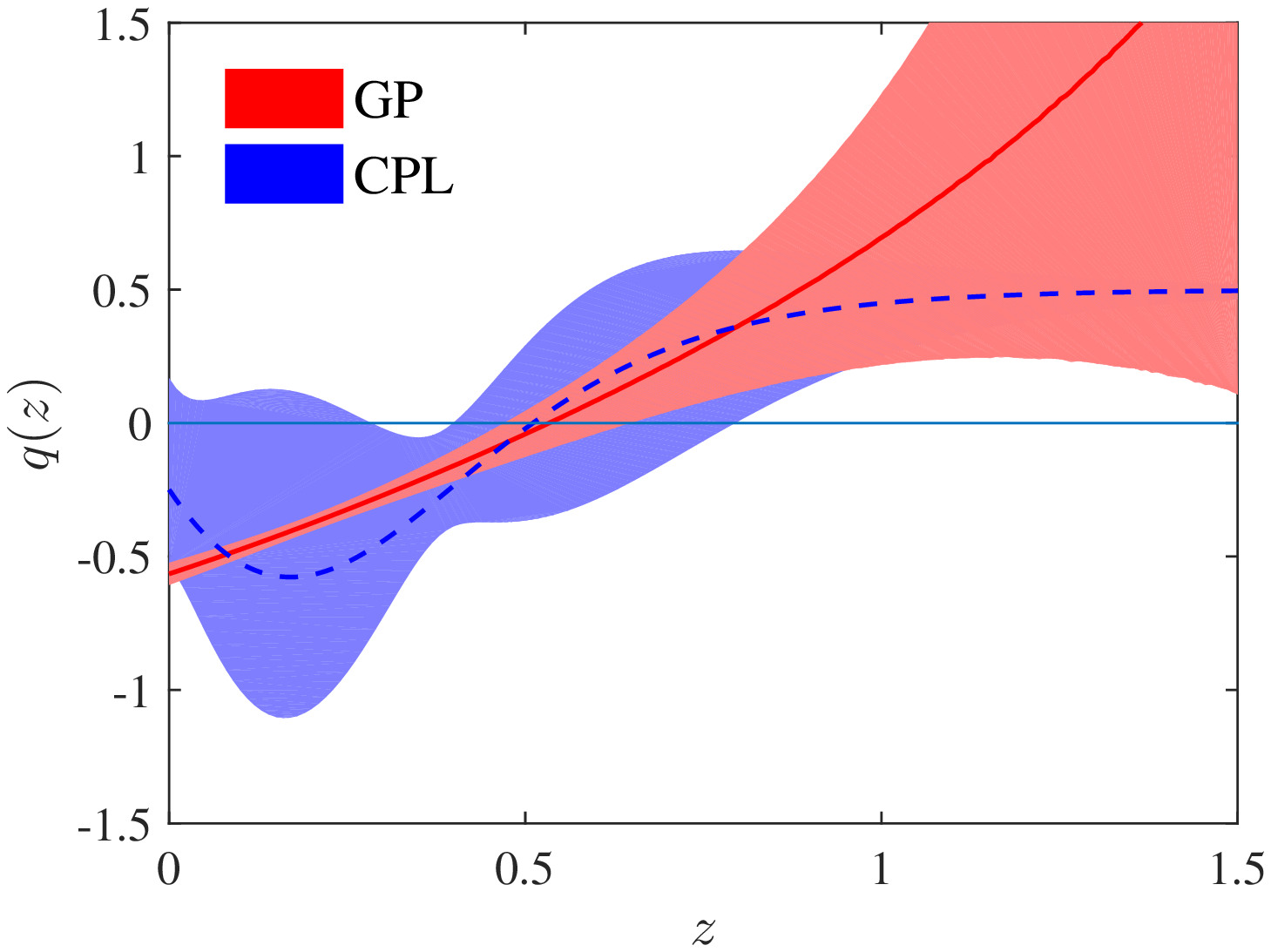}
\caption{Reconstruction of $D$ and its derivatives in the CPL model and GP method using Union2 compilation. The shaded regions are their errors within 68\% C.L. The solid and dashed lines are mean values of reconstruction in the GP method and CPL model, respectively.}  \label{comparison}
\end{figure*}

In previous work \cite{li2011examining,castro2012scaling}, Union2 compilation was found being able to favor the possible slowing down of cosmic acceleration under the CPL model. Instead, GP reconstruction in Fig. \ref{qunion2}  is against such a result. In this section, we try to explore the reason of this tension.

According to the definition of Eq. \eqref{D:define}, dimensionless distance $D(z)$ can be obtained from the observational luminosity distance, without any assumption on $H(z)$ in Eq. \eqref{dL:define}. However, for the parametrization constraint, a hypothesis on $H(z)$ is inevitably imposed. For the CPL parametrization, distance $D(z)$ in Eq. \eqref{D:define} should obey the following prior on Hubble parameter
\be
\frac{H^2(z)}{H_0^2} =   \Omega_m (1+z)^3 + (1-\Omega_m)(1+z)^{3(1+w_0+w_a) } \exp \left( \frac{-3 w_a z}{1+z} \right) .
\ee
Here we adopt the estimation by Union2 data in Ref. \cite{su2011figure}, namely, matter density $\Omega_m=0.419^{+0.090}_{-0.028}$, parameters $w_0=-0.86^{+0.46}_{-0.32}$ and $w_a=-5.51^{+6.99}_{-8.79}$. In Fig. \ref{comparison}, we plot the reconstructions in CPL model and GP method. First panel shows that the CPL model fit well with the observational data. For higher order derivative of $D$, the reconstructions in CPL model are obviously against the GP reconstruction, especially for redshift about $z<1$. In cosmology, there are so many dark energy models fitting well with the supernova data. CPL parametrization is one of them. However, not all models that fit well with the observational data are the essential description of the cosmic expansion history. For higher order derivative of distance $D$, CPL gradually presents an inconsistent result with the true cosmic expansion. Therefore, from the comparison in Fig.  \ref{comparison}, we think that slowing down of acceleration in the CPL model is only a ``mirage".

\section{Conclusion and discussion}
\label{conclusion}

Possible slowing down of cosmic acceleration has attracted great attention. However, most investigations on deceleration parameter were carried out in specific models. In this work, we studied it using GP technique with the Union2, Union2.1 and GRB data, and OHD. Our model-independent analysis suggested that slowing down of cosmic acceleration cannot be obtained by current data within 95\% C.L., in considering the influence of spatial curvature and Hubble constant.

One improvement of our approach concerns the GP method on deceleration parameter, allowing us to break the limitation from specific model. Although many parameterizations \cite{wang2016comprehensive} were considered, there still exists some tensions because of the drive of  model-dependence. On the one hand, they were busy balancing the tension between different models. As discovered in Ref.  \cite{li2010probing,li2011examining}, the same data combination may present different judgements for diverse parametrization. On the other hand, the same model also receive different comments from diverse observational data \cite{magana2014cosmic,wang2016comprehensive}. Tensions made the relevant study stands unstably. Estimations by the novel GP technique are more stable and objective, in contrast to the model parametrization.

Another potential difference of our work is the extension of data types. In previous analysis luminosity distance was the most popular \cite{li2010probing,magana2014cosmic,wang2016comprehensive}. First, we develop the use of luminosity distance from sole supernova data to GRB with higher redshift. Of course, credibility of current GRB data indeed needs to be improved. It indicates that we still lack the knowledge of GRB-cosmology. Second, we promote the use of distance measurement to dynamical $H(z)$ measurement.  Moreover, we compile the latest OHD combination, and split them into two teams, to respectively explore their contribution to the reconstruction.

In addition, our consideration on spatial curvature and Hubble constant also presents a full complement to previous study on this subject. A similar study was found in earlier literature \cite{seikel2012reconstruction}, but we share the aforementioned benefits to a more general case and larger volume data.

One point we should highlight is our analysis on the tension in Section \ref{Union2_discussion}. In contrast to previous CPL model, Union2 data in the GP approach do not favor a slowing down of acceleration. It tells us that slowing down of acceleration in the CPL-like model is only a ``mirage". Although these parameterizations fits well with the observational data, the tension can be revealed by high order derivative of distance $D$. Instead, GP method is able to faithfully model the cosmic expansion history.

In the future, we will try to promote the GP technique for more data, such as the BAO, CMB and other types of data, thereby allowing us to determine the cosmic acceleration with higher accuracy, and to gain richer insight on the physical origins of possible slowing down. Following our present work, we also would like to further investigate cosmic acceleration by revealing the relation between kinematics and dark energy.


\section*{Acknowledgments}

We thank the anonymous referee whose suggestions greatly helped us improve this paper. We thank Vinicius C. Busti, Sahba Yahya and Jing-Zhao Qi for the help on Gaussian processes. J.-Q. Xia is supported by the National Youth Thousand Talents Program and the National Science Foundation of China under grant No. 11422323. M.-J. Zhang is funded by China Postdoctoral Science Foundation under grant No. 2015M581173. The research is also supported by the Strategic Priority Research Program ¡°The Emergence of Cosmological Structures¡± of the Chinese Academy of Sciences, grant No. XDB09000000.

\bibliography{evolution}
\end{document}